\documentclass[prx,aps,superscriptaddress,nofootinbib,twocolumn]{revtex4-1}

\usepackage{mathrsfs}
\usepackage{amsfonts}
\usepackage{amssymb}
\usepackage{amsmath}
\usepackage{bm}
\usepackage{graphicx}
\usepackage[usenames,dvipsnames]{color}
\usepackage[colorlinks=true,citecolor=blue,linkcolor=magenta]{hyperref}
\usepackage{ulem}
\usepackage{lmodern}
\usepackage{ulem}

\newcommand{\figpath}{./}

\newcommand{\Tr}{\mathrm{Tr}}
\newcommand{\norm}[1]{\| #1 \|}
\newcommand{\abs}[1]{| #1 |}

\newcommand{\ket}[1]{\vert{ #1 }\rangle}

\newcommand{\ketbra}[2]{\vert #1 \rangle \langle #2 \vert}

\newcommand{\mean}[1]{\langle #1 \rangle}

\newcommand{\rket}[1]{\vert{ #1 }\rangle\rangle}
\newcommand{\rbra}[1]{\langle\langle{ #1 }\vert}
\newcommand{\rbraket}[2]{\langle\langle #1 \vert #2 \rangle\rangle}
\newcommand{\OO}{\mathcal{O}}
\newcommand{\BB}{\mathcal{B}}
\newcommand{\EE}{\mathcal{E}}
\newcommand{\NN}{\mathcal{N}}
\newcommand{\TT}{\mathcal{T}}
\newcommand{\II}{\mathcal{I}}
\newcommand{\MM}{\mathcal{M}}
\newcommand{\GG}{\mathcal{G}}

\newcommand{\RED}[1]{{\color{black} #1}}

\begin{document}

\title{Practical Quantum Error Mitigation for Near-Future Applications}

\author{Suguru Endo}

\author{Simon C. Benjamin}

\affiliation{Department of Materials, University of Oxford, Oxford OX1 3PH}

\author{Ying Li}

\affiliation{Graduate School of China Academy of Engineering Physics, Beijing 100193, China}
\affiliation{Department of Materials, University of Oxford, Oxford OX1 3PH}

\begin{abstract}
It is vital to minimise the impact of errors for near-future quantum devices that will lack the resources for full fault tolerance. Two quantum error mitigation (QEM) techniques have been introduced recently, namely error extrapolation~\cite{Li2017,Temme2017} and quasi-probability decomposition~\cite{Temme2017}. To enable practical implementation of these ideas, here we account for  the inevitable imperfections in the experimentalist's knowledge of the error model itself. We describe a protocol for systematically measuring the effect of errors so as to design efficient QEM circuits. We find that the effect of localised Markovian errors can be fully eliminated by inserting or replacing some gates with certain single-qubit Clifford gates and measurements. Finally, having introduced an exponential variant of the extrapolation method we contrast the QEM techniques using exact numerical simulation of up to 19 qubits in the context of a `{\small SWAP} test' circuit. Our optimised methods dramatically reduce the circuit's output error without increasing the qubit count or time requirements.
\end{abstract}

\maketitle

\section{Introduction}

Controlling noise in quantum systems is crucial for the development of practical technologies. Such noise can occur due to unwanted interactions of a passive qubit with the environment, or due to imperfections in the use of circuit elements that compose the algorithm (qubit initialisation, gates, and measurement). In all cases the result is errors occurring at the level of physical qubits. The theory of quantum fault tolerance (QFT) reveals that the introduction of logical qubits, composed of numerous physical qubits, can allow one to detect and correct errors at the physical level; however this capacity comes at an enormous multiplicative cost in resources. A recent estimate suggests that a Shor algorithm operating on a few thousand logical qubits would require several million physical qubits~\cite{Joe2016}. While it is encouraging to know that such techniques exist, hardware on this scale is probably at least a decade away. The timely (indeed, urgent) question is, to what extent can we control the impact of errors in computing devices that are too small to support full QFT?

It may prove to be the case that deep quantum algorithms, such as Shor's factoring algorithm and Grover's search algorithm, cannot be successfully executed on classically-intractable problems without the support of QFT. However, fortunately there are other algorithms of potential practical significance that focus on shallow circuits, with the output typically being fed into a classical supervising algorithm so as to form a hybrid system. Such approaches have been proposed for the simulation to aid discovery in chemistry and materials science; see Refs.~\cite{Li2017, Peruzzo2014, Wecker2015, McClean2016, Bauer2016, Kreula2016, Kreula2016EPJ, Shen2017, OMalley2016, Kandala2017} for examples. Hybrid systems may be capable of yielding significant results, surpassing conventional computers, even when finite error rates are present \RED{because of their error resilience~\cite{McClean2017, Colless2018}}. In order to achieve this it is desirable to suppress or mitigate errors to the greatest extent possible while keeping the qubit count ideally unchanged, or \RED{increasing only modestly compared to the high cost of full QFT}. 

Recently two techniques were introduced for quantum error mitigation (QEM) in generic hybrid quantum algorithms where the expected value of an observable -- say, a $z$-basis measurement of a given qubit -- is the quantity of interest. The goal is to estimate the value that this observable would take given an error-free circuit, despite the reality that the real experimental system cannot perform operations with less than a certain error rate. Ref.~\cite{Li2017} introduced a hybrid algorithm simulating quantum dynamics, which featured an active error minimisation technique involving extrapolation. The experimentalist would execute the circuit with all errors at their minimum achievable severity, obtain the expected value of the observable, and then repeat the exercise having deliberately increased the physical error rate (or having applied additional {quantum gates} to achieve the same effect). By noting the effect of the increased errors on the observable, the experimentalist would be able to make an extrapolated estimate of the zero-error value, presuming that the error sources had scaled proportionately. The technique was found to be very advantageous in the numerical simulations of few-qubit experiments presented in that paper (see e.g.~Fig. 5 in Ref.~\cite{Li2017}). 

A paper that appeared online at almost the same time was Ref.~\cite{Temme2017} by the IBM-based team of Temme, Bravyi and Gambetta. This paper presented a comprehensive analysis of the extrapolation technique, which the authors had independently conceived, and moreover it introduced a second technique with using (what we will call) a `quasi-probability' formalism. The authors explained that by replacing operations in the quantum circuit and assigning parity $\pm1$ to each operation following a certain probability distribution dependent on the noise, an experimentalist can obtain the unbiased estimator, at the cost of an increase in the variance. Their method was shown to be applicable to specific noise types including homogeneous depolarizing errors and damping errors. The authors found both methods to be promising in few-qubit numerical simulations (see e.g. Fig. 2 in Ref.~\cite{Temme2017}).
 
As exciting as these studies were, open questions remained to be answered before these two techniques could be considered to be fully practical. First, both techniques rely on the full knowledge of the error model, whereas an experimentalist will have imperfect knowledge and the real noise will generally differ from the canonical types considered in these first papers. Second, we need an explicit method to derive the QEM circuits, i.e. a specification of how to algorithmically increase the error rate in the error extrapolation or how to sample circuits in the quasi-probability decomposition. In this paper, we solve these two problems. We find that gate set tomography (GST)~\cite{Merkel2013, Greenbaum2015} provides sufficient information to enable full elimination of {the impact of} localised Markovian errors. As with other process tomography protocols, GST cannot determine the exact physical error model {due to noise associated with state preparation and measurement}. However, we determine that {preparation and measurement noise} in GST is not harmful to the overall QEM approach. We also find that single-qubit Clifford gates and measurements are universal in computing expected values. Each quantum operation is a linear map, and single-qubit Clifford gates and measurements yield a complete set of linearly-independent maps (quantum operations). Therefore, any error can be simulated or subtracted by decomposing the error using the complete operation set, which is the standard linear decomposition. We prove that, by combining GST and the complete set decomposition, any localised and Markovian errors in the quantum computer can be systemically mitigated, so that the error in the final computational output is only due to {unbiased} {statistical fluctuation}.

For the quasi-probability method, we provide an upper bound of the cost in QEM, and we describe the utility of `twirling' operations~\cite{Knill2004, Wallman2015, OGorman2016} in minimising this cost. For the extrapolation method, which is a relatively straightforward technique, our optimisation is to observe that typically for the classes of noise most common in experiments it is appropriate to assume that the expected value of the observable will decay exponentially with the severity of the circuit noise. Adopting this underlying assumption, rather than a polynomial (e.g. linear) fit, proves to be quite advantageous. 

Having thus optimised both the quasi-probability and the extrapolation techniques, we make a series of numerical simulations to study their efficacy. We opt for a specific circuit, a realisation of the `{\small SWAP} test' that is often employed in quantum algorithms as a means for estimating the similarity of quantum states~\cite{Ekert2002,shallowswap}. Our {\small SWAP} test operates on $2n+1$ qubits, and we simulate a total of $15$ qubits over a comprehensive set of cases as well as $19$ qubits for specific cases. We numerically simulate the actions of the experimentalist, who must perform many circuit trials in order to make a single estimate of the observable (we choose $10^4$ trials). But in order to evaluate our QEM techniques we must then repeat this entire process to determine the distribution of values that the experimentalist might obtain. We perform at least $10^3$ repetitions so that the distribution becomes clear, thus at least $10^7$ individual numerical experiments are performed for each of the curves that we presently report.

\section{Error mitigation}

\begin{figure*}[tbp]
\centering
\includegraphics[width=1\linewidth]{\figpath 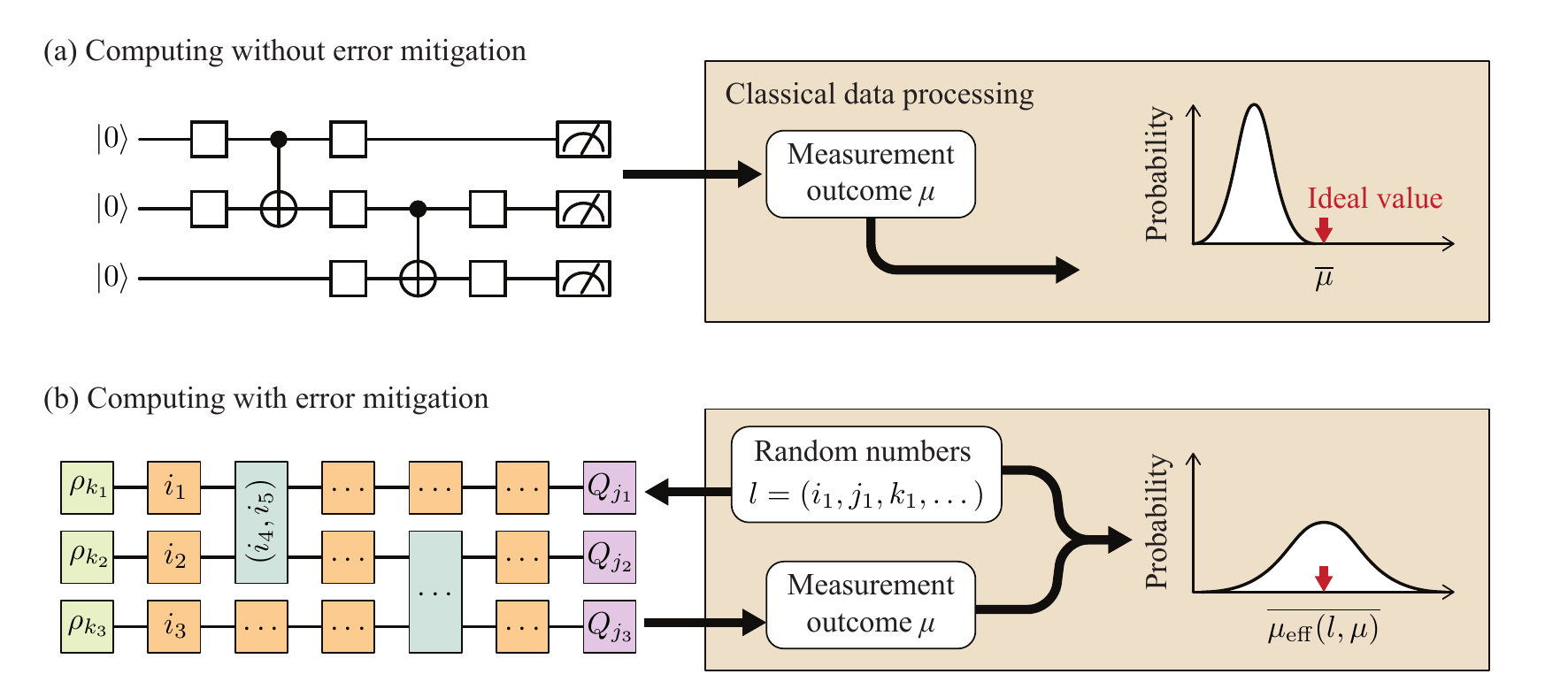}
\caption{
Quantum computing of the expected value of an observable (a) without quantum error mitigation (QEM) and (b) with QEM. In QEM circuit (b), each operation (including the memory operation) in the original circuit (a) is replaced by an operation depending on the corresponding random numbers [see Fig.~\ref{fig:circuit}(a)].
}
\label{fig:protocol}
\end{figure*}

In this paper, we focus on computing the expected value of an observable in a state (the final state of a quantum circuit) using a quantum computer. It is typical of a number of quantum algorithms and subroutines that the desired output is the expected value of a qubit or qubits -- the {\small SWAP}-test~\cite{Ekert2002,shallowswap} itself, which is a component of algorithms including the recently-introduced auto encoder~\cite{AspuruGuzikAutoEnc}, and several proposed hybrid algorithms for simulating chemical or materials systems~\cite{Li2017, Peruzzo2014, Wecker2015, McClean2016, Bauer2016, Kreula2016, Kreula2016EPJ}.

Without using QEM as shown in Fig.~\ref{fig:protocol}(a), the quantum circuit is repeated for many times, and the measurement outcome $\mu$ of each time is collected. Then, we can calculate the average $\overline{\mu}$ as our best estimate of the expected value. {Given that the number of repetitions is finite, the value of $\overline{\mu}$ is a random variable with an associated distribution}. Because the implementation of the quantum circuit is imperfect, it is likely that the distribution of $\overline{\mu}$ is not even centered at the ideal value, i.e.~the exact expected value when the quantum circuit is perfectly implemented without error.

When we use QEM as shown in Fig.~\ref{fig:protocol}(b), instead of the original quantum circuit, we implement a set of modified circuits. The scheme depicted in the figure is relevant to the quasi-probability method for QEM, but can also apply to the extrapolation method as a means to deliberately boost errors. Each modified circuit is determined by a set of random numbers $l$. The distribution of random numbers, i.e.~modified circuits, depends on the error model, which is measured using GST before the quantum computing. In each run of the quantum experiment, firstly the random number set $l$ is generated, then depending on $l$ a specific circuit is implemented, and finally the measurement outcome $\mu$ is collected. Rather than calculating the average $\overline{\mu}$, we use both $l$ and $\mu$ to calculate the average of an effective outcome $\mu_{\rm eff}(l,\mu)$, which will be given explicitly later. If QEM is successful, the distribution of $\overline{\mu_{\rm eff}(l,\mu)}$ is centered at the ideal value, but the distribution is wider than $\overline{\mu}$. Thus only error due to the statistical fluctuation remains, although it is amplified. By repeating the quantum experiment enough times, we can obtain an accurate computing result of the expected value.

In Sec.~\ref{sec:sampling}, we explicitly give the effective outcome $\overline{\mu_{\rm eff}(l,\mu)}$. Modified circuits and their distribution are given in Sec.~\ref{sec:decomposition}.

\section{Notation for states, operators and operations}

We use the notation commonly used in quantum tomography (e.g.~in Refs.~\cite{Merkel2013, Greenbaum2015}).

In quantum theory, a quantum state is usually represented by a density matrix $\rho$, and an observable is represented by a Hermitian operator $Q$. The expected value of the observable quantity in the state is $\mean{Q} = \Tr(Q\rho)$. An operation is a map on the space of states, $\OO(\rho) = \sum_k E_k \rho E_k^\dag$, expressed in the Kraus form.

Because an operation is a linear map, we can always express the operation $\OO$ as a square matrix, e.g.~using the Pauli transfer matrix representation, acting on the state expressed as a column vector $\rket{\rho}$. Similarly, an observable can be expressed as a row vector $\rbra{Q}$, and the expected value is $\mean{Q} = \rbraket{Q}{\rho}$. Throughout this paper, we use the Pauli transfer matrix representation, and see Appendix~\ref{app:PTM} for details. In quantum tomography, usually we focus on observables that are POVM operators, which is not necessary here.

In the Pauli transfer matrix representation, vectors representing states or observables and matrices representing operations are all real. For $n$ qubits, vectors and matrices are $4^n$-dimensional. The expected value of the observable $Q$ in the state $\rho$ going through a sequence of operations $\OO_1, \ldots, \OO_N$ reads as follows:
\[
\Tr[Q\OO_N \circ \cdots \circ \OO_1(\rho)] = \rbra{Q} \OO_N \cdots \OO_1 \rket{\rho}.
\]

\section{Quantum computing by sampling circuits}
\label{sec:sampling}

We suppose that the initial state is $\rho^{(0)}$, which goes through a sequence of operations $\OO_1^{(0)}, \ldots, \OO_N^{(0)}$, and in the final state the observable $Q^{(0)}$ is measured. Each time the experimentalist implements this circuit, the measurement returns an eigenvalue of $Q^{(0)}$, and the probability distribution of eigenstates is determined by the final state. By repeating such a circuit for many times, she can estimate the expected value $\mean{Q^{(0)}} = \rbra{Q^{(0)}} \OO_{\rm tot.}^{(0)} \rket{\rho^{(0)}} = {\rm E}[\mu^{(0)}]$, where $\OO_{\rm tot.}^{(0)} = \OO_N^{(0)} \cdots \OO_1^{(0)}$, and $\mu^{(0)}$ is the measurement outcome. Generally in this paper we will use the superscript $0$ to denote the ideal noise-free realisation of {a state,} operation or observable quantity.

In the case that the quantum computation has errors, the actual initial state is $\rho$, actual operations are $\OO_1, \ldots, \OO_N$, and the actually measured observable is $Q$. As a result, the estimation of the expected value converges to $\mean{Q} = \rbra{Q} \OO_{\rm tot.} \rket{\rho}$ rather than $\mean{Q^{(0)}}$. Here, $\OO_{\rm tot.} = \OO_N \cdots \OO_1$, and we have assumed that errors are Markovian.

The central idea introduced by the IBM team in Ref.~\cite{Temme2017} is that one can exactly compensate for the effect of errors by sampling from a set of (real, error-burdened) circuits, each labelled $\OO_{\rm tot.}^{(l)}$ for $l=1,2...$, provided that their outputs satisfy 
\[
\mean{Q^{(0)}} = \sum_l q_l \rbra{Q^{(l)}} \OO_{\rm tot.}^{(l)} \rket{\rho^{(l)}}.
\] 
Ref.~\cite{Temme2017} describes how the real numbers $\{ q_l \}$ which represent quasi-probabilities can be efficiently derived given {specific error models, assuming that the experimentalist has full knowledge of the model}. Note that each $\OO_{\rm tot.}^{(l)}$ denotes the total operation composed by a sequence of operations in the $l^\text{th}$ circuit.

We can use the Monte Carlo method to compute $\mean{Q^{(0)}}$. We note that $\rbra{Q^{(l)}} \OO_{\rm tot.}^{(l)} \rket{\rho^{(l)}} = {\rm E}[\mu^{(l)}]$, where $\mu^{(l)}$ is the measurement outcome in the $l^\text{th}$ circuit. Then $\mean{Q^{(0)}} = \sum_l \abs{q_l} {\rm E}[{\rm sgn}(q_l)\mu^{(l)}]$. To compute $\mean{Q^{(0)}}$, we randomly choose a circuit to implement, and the $l^\text{th}$ circuit is chosen with the probability $p_l = \abs{q_l}/C$, where $C = \sum_l \abs{q_l}$. Then, the computing result is given by the expected value of effective measurement outcomes, i.e.~$\mean{Q^{(0)}} = C {\rm E}[\mu_{\rm eff}]$, where the effective outcome is $\mu_{\rm eff} = {\rm sgn}(q_l)\mu^{(l)}$ if the $l^\text{th}$ circuit is chosen to be implemented, and $\mu^{(l)}$ is the outcome directly obtained in the $l^\text{th}$ circuit.

\section{Per-operation error correction}
\label{sec:per-operation}

We can correct errors in each operation using the quasi-probability method, which will be the primary focus for the following several sections. We suppose that we have a set of initial states satisfying $\rket{\rho^{(0)}} = \sum_{l_{\rm in}} q_{l_{\rm in}}^{[{\rm in}]} \rket{\rho^{(l_{\rm in})}}$, and a set of operations satisfying $\OO_i^{(0)} = \sum_{l_i} q_{l_i}^{[i]} \OO_i^{(l_i)}$ for each error-free operation $\OO_i^{(0)}$, and a set of observables satisfying $\rbra{Q^{(0)}} = \sum_{l_{\rm out}} q_{l_{\rm out}}^{[\rm out]} \rbra{Q^{({l_{\rm out}})}}$. Then, computing with error mitigation can be expressed as
\begin{eqnarray}
\mean{Q^{(0)}} &=& \sum_{l_{\rm in}} \sum_{l_1} \cdots \sum_{l_N} \sum_{l_{\rm out}}
q_{l_{\rm in}}^{[{\rm in}]} q_{l_1}^{[1]} \cdots q_{l_N}^{[N]} q_{l_{\rm out}}^{[\rm out]} \notag \\
&&\times \rbra{Q^{({l_{\rm out}})}} \OO_N^{(l_N)} \cdots \OO_1^{(l_1)} \rket{\rho^{(l_{\rm in})}}.
\end{eqnarray}
When we sample circuits to compute $\mean{Q^{(0)}} = C {\rm E}[\mu_{\rm eff}]$, the initial state is $\rket{\rho^{(l_{\rm in})}}$ with  probability $p_{l_{\rm in}}^{[{\rm in}]} = q_{l_{\rm in}}^{[{\rm in}]}/C_{\rm in}$, the $i^\text{th}$ operation is $\OO_i^{(l_i)}$ with  probability $p_{l_i}^{[i]} = \abs{q_{l_i}^{[i]}} / C_i$, and the observable is $\rbra{Q^{({l_{\rm out}})}}$ with  probability $p_{l_{\rm out}}^{[{\rm out}]} = q_{l_{\rm out}}^{[{\rm out}]}/C_{\rm out}$. Here, $C_\alpha = \sum_{l_\alpha} \abs{q_{l_\alpha}^{[\alpha]}}$, and $C = C_{\rm in} C_1 \cdots C_N C_{\rm out}$ accordingly. To calculate $\mu_{\rm eff}$, we use ${\rm sgn}(q_{l_{\rm in}}^{[{\rm in}]} \cdots q_{l_{\rm out}}^{[\rm out]}) = {\rm sgn}(q_{l_{\rm in}}^{[{\rm in}]}) \cdots {\rm sgn}(q_{l_{\rm out}}^{[\rm out]})$.

\section{Variance amplification in quasi-probability decomposition}

The presence of quasi-probabilities taking negative values amplifies the variance of the expected value of the observable. We consider the case that $Q^{(l)}$ is a Pauli operator (maybe with error) and the measurement reports two kinds of outcomes denoted by $\pm 1$, respectively. In this case, the distribution is binomial. The standard deviation of \RED{the average of outcomes in} the Monte Carlo calculation is $\sigma = C \sqrt{(1-{\rm E}[\mu_{\rm eff}]^2)/N_{\rm r}} \leq C/\sqrt{N_{\rm r}}$. Here, $N_{\rm r}$ is the total number of samples, i.e.~{the total number of circuits of all kinds which the experimentalist performs is $N_{\rm r}$}. We compare this to the error-free computing, i.e.~the ideal original circuit $\rbra{Q^{(0)}} \OO_{\rm tot.}^{(0)} \rket{\rho^{(0)}}$ is repeated for $N_{\rm r}^{(0)}$ times to estimate $\mean{Q^{(0)}}$. For the error-free computing, the standard deviation is given by $\sigma^{(0)} = \sqrt{(1-{\rm E}[\mu^{(0)}]^2)/N_{\rm r}^{(0)}}$. Therefore, to achieve the same accuracy, i.e.~$\sigma = \sigma^{(0)}$, the error-mitigated computation needs $N_{\rm r}/N_{\rm r}^{(0)} = (C^2- \mean{Q^{(0)}}^2)/(1-\mean{Q^{(0)}}^2)$ times more samples than the error-free computation. Here, we have used the fact that the error-mitigated computation and the error-free computation should converge to the same value of $\mean{Q^{(0)}}$, i.e.~${\rm E}[\mu^{(0)}] = C {\rm E}[\mu_{\rm eff}]$.

In order to limit the standard deviation to be $\sigma \sim \epsilon$, we can choose $N_{\rm r} \sim (C/\epsilon)^2$. Therefore, if the factor $C$ is larger, the computing takes longer.

Because $C = C_{\rm in} C_1 \cdots C_N C_{\rm out}$ if errors are corrected for each operation, we call $C_\alpha - 1$ the cost for mitigating error in the corresponding operation. The overall cost therefore increases with the number of operations, thus it is important to reduce the operation number, e.g.~in quantum computer with qubits fully connected~\cite{Song2017}, operations for communication are not required, which may significantly reduce the cost.

\section{Universal operation set}

\begin{table}[tbp]
\begin{center}
\begin{tabular}{|c|l|}
\hline
1 & ~~$[\openone]$ (no operation) \\
\hline
2 & ~~$[\sigma^{\rm x}] = [R_{\rm x}]^2$ \\
\hline
3 & ~~$[\sigma^{\rm y}] = [R_{\rm x}]^2[R_{\rm z}]^2$ \\
\hline
4 & ~~$[\sigma^{\rm z}] = [R_{\rm z}]^2$ \\
\hline
5 & ~~$[R_{\rm x}] = [\frac{1}{\sqrt{2}}(\openone +i \sigma^{\rm x})] = [H][S]^3[H]$ \\
\hline
6 & ~~$[R_{\rm y}] = [\frac{1}{\sqrt{2}}(\openone +i \sigma^{\rm y})] = [R_{\rm z}]^3[R_{\rm x}][R_{\rm z}]$ \\
\hline
7 & ~~$[R_{\rm z}] = [\frac{1}{\sqrt{2}}(\openone +i \sigma^{\rm z})] = [S]^3$ \\
\hline
8 & ~~$[R_{\rm yz}] = [\frac{1}{\sqrt{2}}(\sigma^{\rm y} + \sigma^{\rm z})] = [R_{\rm x}][R_{\rm z}]^2$ \\
\hline
9 & ~~$[R_{\rm zx}] = [\frac{1}{\sqrt{2}}(\sigma^{\rm z} + \sigma^{\rm x})] = [R_{\rm z}][R_{\rm x}][R_{\rm z}]$ \\
\hline
10 & ~~$[R_{\rm xy}] = [\frac{1}{\sqrt{2}}(\sigma^{\rm x} + \sigma^{\rm y})] = [R_{\rm x}]^2[R_{\rm z}]$ \\
\hline
11 & ~~$[\pi_{\rm x}] = [\frac{1}{2}(\openone + \sigma^{\rm x})] = [R_{\rm z}]^3[R_{\rm x}]^3[\pi][R_{\rm x}][R_{\rm z}]$ \\
\hline
12 & ~~$[\pi_{\rm y}] = [\frac{1}{2}(\openone + \sigma^{\rm y})] = [R_{\rm x}][\pi][R_{\rm x}]^3$ \\
\hline
13 & ~~$[\pi_{\rm z}] = [\frac{1}{2}(\openone + \sigma^{\rm z})] = [\pi]$ \\
\hline
14 & ~~$[\pi_{\rm yz}] = [\frac{1}{2}(\sigma^{\rm y} +i \sigma^{\rm z})] = [R_{\rm z}]^3[R_{\rm x}]^3[\pi][R_{\rm x}]^3[R_{\rm z}]$ \\
\hline
15 & ~~$[\pi_{\rm zx}] = [\frac{1}{2}(\sigma^{\rm z} +i \sigma^{\rm x})] = [R_{\rm x}][\pi][R_{\rm x}]^3[R_{\rm z}]^2$ \\
\hline
16 & ~~$[\pi_{\rm xy}] = [\frac{1}{2}(\sigma^{\rm x} +i \sigma^{\rm y})] = [\pi][R_{\rm x}]^2$ \\
\hline
\end{tabular}
\end{center}
\caption{
Sixteen basis operations. Gates $[R_{\rm x}]$ and $[R_{\rm y}]$ can be derived from $[H]$ and $[S]$, and other operations can be derived from $[\pi]$, $[R_{\rm x}]$ and $[R_{\rm y}]$.
}
\label{tab:bases}
\end{table}

The set of operations including measurement and single-qubit Clifford gates is universal in computing expected values of observables. The relevant measurement operation reads $[\pi] = [\frac{1}{2}(\openone + \sigma^{\rm z})]$, which projects a qubit to the state $\ket{0}$ . Here, $[U](\rho) = U\rho U^\dag$ denotes a superoperator. Such a non-destructive measurement can be realised using a destructive measurement followed by initialising the qubit in the state $\ket{0}$. Single-qubit Clifford gates include the Hadamard gate $[H] = [\frac{1}{\sqrt{2}}(\sigma^{\rm x} + \sigma^{\rm z})]$, the phase gate $[S] = [\frac{1}{\sqrt{2}}(\openone -i \sigma^{\rm z})]$ and all other \RED{single-qubit Clifford} gates can be derived from these two.

The measurement superoperator $[\pi]$ also means post-selection, i.e.~if the outcome of the measurement corresponding to $[\pi]$ (which is not the final measurement on the observable $Q^{(l)}$) is $\ket{1}$ in a trial, the value of the observable $Q^{(l)}$ is noted as $\mu^{(l)} = 0$, but the trial is counted in the total number of samples in the Monte Carlo calculation. If $Q^{(l)}$ has two values $\pm 1$, we can estimate the value of $\rbra{Q^{(l)}} \OO_{\rm tot.}^{(l)} \rket{\rho^{(l)}}$ by calculating $(N_{+1}^{(l)} - N_{-1}^{(l)})/(N_{0}^{(l)} + N_{+1}^{(l)} + N_{-1}^{(l)})$. Here, we have supposed that the circuit is implemented for total $N_{0}^{(l)} + N_{+1}^{(l)} + N_{-1}^{(l)}$ times; for $N_{0}^{(l)}$ times the circuit does not pass post-selections (i.e.~$\mu^{(l)} = 0$), and for $N_{\pm 1}^{(l)}$ times the circuit passes all post-selections and reports $Q^{(l)} = \pm 1$ (i.e.~$\mu^{(l)} = \pm 1$). It is the same when we compute $\mean{Q^{(0)}}$ using the Monte Carlo method. If the effect outcome is $\mu_{\rm eff} = 0$ with the probability $P_0$, then the standard deviation of the Monte Carlo calculation becomes $\sigma = C \sqrt{[(1-P_0)^2-{\rm E}[\mu]^2]/(1-P_0)N_{\rm r}} \leq C\sqrt{(1-P_0)/N_{\rm r}}$.

In Table~\ref{tab:bases}, we list sixteen linearly independent operations that can be derived from the minimum universal operation set $\{ [\pi], [H], [S] \}$. In the following, we use $\{\BB_i^{(0)} \vert i=1,\ldots,16\}$ to denote these sixteen operations. Because they are linearly independent, any single-qubit operation $\OO$, which is a $4\times 4$ real matrix, can be expressed as a linear combination of sixteen basis operations, i.e.~$\OO = \sum_{i=1}^{16} q_i \BB_i^{(0)}$. Similarly, multi-qubit operations can be expressed as a linear combination of tensor products of basis operations. Using the quasi-probability method, any computation of expected values of observables can be realised using this operation set.


\RED{Note that these basis operations are universal, as one can verify by constructing a non-Clifford gate or an entangling gate: We can decompose $T$ gate using our basis operations as $[T] = \frac{1}{2}[\openone] - \frac{\sqrt{2}-1}{2}[\sigma^{\rm z}] + \frac{\sqrt{2}}{2}[R_{\rm z}^3]$ (see Appendix~\ref{app:CNOT} for controlled-{\small NOT} as a second example). However this construction would not be used in practice -- it is not an efficient means to actually implement a desired $T$ in the basic circuit, since the corresponding cost $C = \sqrt{2}$ would imply and unacceptably steep exponential in the time overhead, as one would expect from e.g. Refs.~\cite{Bravyi2016, Bravyi2016PRX, Howard2017}. Instead we rely on the assumption that the experimental system can directly implement a universal set of gates (including entangling and non-Clifford gates) with a reasonably high fidelity. Then rather than fully synthesising any of the basic gates using our basis, we need only compensate for slight imperfections. The cost for doing so, for each imperfect gate, is then $\sim C = 1+\delta$ as we presently discuss.}

Having obtained the complete operation set in Table~\ref{tab:bases} we can use it in deriving the protocol that will compensate for errors. In this paper, we focus on the case that errors are localised: An (error-free) operation that is applied on a set of qubits $S$ is a $4^{\abs{S}}$-dimensional real matrix, then the corresponding operation in real (i.e. error-burdened) $\OO$ is also a $4^{\abs{S}}$-dimensional real matrix acting on the same set of qubits. The overall operation on the entire system can be expressed as $\openone_{\bar{S}}\otimes \OO$, where $\openone_{\bar{S}}$ is the identity acting on all other qubits. It is similar for the initialisation and measurement. If each qubit is initialised individually, the overall initial state is $\bigotimes_m \rket{\rho_m}$, where $\rket{\rho_m}$ is a $2$-dimensional real vector representing the $m^\text{th}$ qubit's initial state. Similarly, individual measurement of qubits implies that the overall measured observable is $\bigotimes_m \rbra{Q_m}$, where $\rbra{Q_m}$ is a $2$-dimensional real vector representing the measured observable for the $m^\text{th}$ qubit. In this case, a single-qubit operation with error can still be expressed using a $4\times 4$ real matrix. We suppose that for a qubit, sixteen basis operations with errors are $\{\BB_i \vert i=1,\ldots,16\}$, which are all $4\times 4$ real matrices. When errors are not significant, these sixteen bases should still be linearly independent, i.e.~the set of basis operations with errors is still universal.

To make this statement more precise, we consider the $16\times 16$ real matrix
\begin{eqnarray}
A = \left[
\begin{array}{ccc}
(\BB_{1})_{\bullet,1} & \cdots & (\BB_{16})_{\bullet,1} \\ 
(\BB_{1})_{\bullet,2} & \cdots & (\BB_{16})_{\bullet,2} \\
(\BB_{1})_{\bullet,3} & \cdots & (\BB_{16})_{\bullet,3} \\
(\BB_{1})_{\bullet,4} & \cdots & (\BB_{16})_{\bullet,4}
\end{array}
\right].
\label{eq:A}
\end{eqnarray}
Here, $(\BB_i)_{\bullet,j}$ denotes the $j^\text{th}$ column of the matrix of the basis operation $\BB_i$. Sixteen basis operations are linearly independent if the matrix $A$ is invertible. We use $\epsilon_{\rm max} = \max\{ \norm{\BB_i - \BB_i^{(0)}}_{\rm max} \vert i=1,\ldots,16 \}$ as the measure of the error severity in basis operations. When $\epsilon_{\rm max} < \frac{1}{32}(13-3\sqrt{17}) \simeq 0.0351$, $A$ is always invertible (see Appendix~\ref{app:BOth}). We remark that even if $\epsilon_{\rm max}$ exceeds the threshold, basis operations are still likely to be linearly independent.

\section{Error mitigation using basis operations}
\label{sec:decomposition}

\begin{figure}[tbp]
\centering
\includegraphics[width=1\linewidth]{\figpath 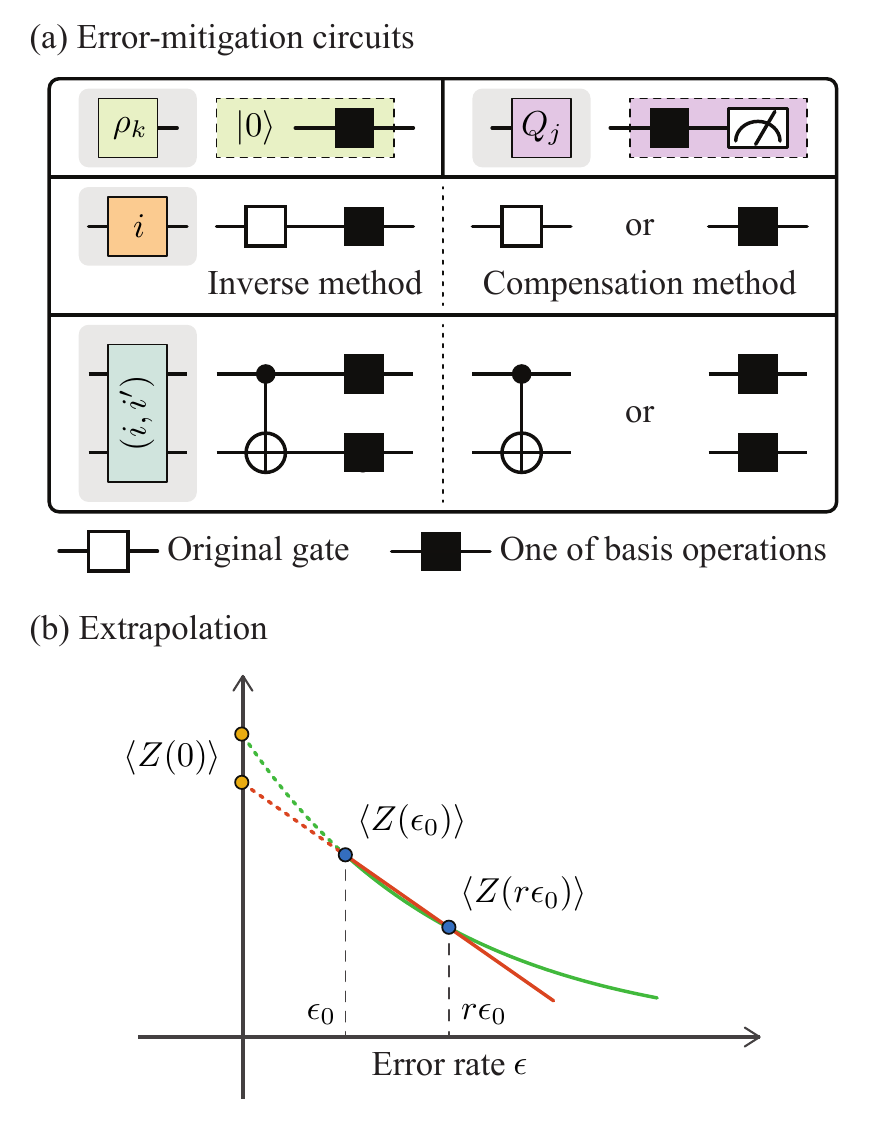}
\caption{
(a) Error-mitigation circuits. The choice of a basis operation is determined by the corresponding random number $i$, $j$ or $k$. Original gate that is identity (memory operation) also has to be error-mitigated, unless memory error is negligible. In the compensation method, ether the original gate or basis operations are applied depending on the random number. (b) The schematic of the linear extrapolation (orange curve) and exponential extrapolation (green curve).
}
\label{fig:circuit}
\end{figure}

Given an operation with error $\OO$, we can use sixteen basis operations to correct the error, i.e.~realise the operation without error $\OO^{(0)}$. There are two ways for correcting the error.

{\bf Compensation method.} The operation $\OO$ is close to $\OO^{(0)}$. Therefore, we can keep the correct component of $\OO$ and only decompose the error component using basis operations. We decompose the operation without error as $\OO^{(0)} = \lambda \OO + \sum_i q_i \BB_i$, where $\lambda$ is an arbitrary real number. If basis operations are linearly independent, the decomposition always exists, and there is only one solution of coefficients $\{q_i\}$ when $\lambda$ is determined.

{\bf Inverse method.} If the matrix $\OO^{(0)}$ is invertible, we can express $\OO$ as $\OO^{(0)}$ followed by a noise operation, i.e.~$\OO = \NN \OO^{(0)}$, where the noise operation $\NN = \OO \OO^{(0)-1}$. In order to correct the error, we can decompose the inverse of the noise as $\NN^{-1} = \OO^{(0)} \OO^{-1} = \sum_i q_i \BB_i$. By applying the inverse of the noise after the operation $\OO$, we can realise the operation without error, i.e.~$\OO^{(0)} = \NN^{-1} \OO = \sum_i q_i \BB_i \OO$. Similar to the compensation method, if basis operations are linearly independent, the decomposition always exists, and there is only one solution of coefficients $\{q_i\}$. However, the inverse method can only be applied if the matrix $\OO$ is invertible.

\RED{For multi-qubit operations, the decomposition is performed using tensor products of basis operations, as described explicitly in Appendix~\ref{app:decom}. Although basis operations are not entangling, we can use basis operations to efficiently mitigate multi-qubit errors and errors that can entangle qubits. As an example, we show how to decompose the controlled-{\small NOT} gate only using basis operations in Appendix~\ref{app:CNOT}, which suffices to imply that any error in the form of the controlled-{\small NOT} gate can be mitigated using basis operations.}

Initialisation and measurement errors can also be corrected using basis operations. Taking first the case of initialisation errors: If $\rket{\rho}$ is the error-burdened initial state, and it is a non-zero vector, we can always find a transformation $\TT$ that satisfies $\rket{\rho^{(0)}} = \TT \rket{\rho}$ where $\rket{\rho^{(0)}}$ is the error-free initial state. Thus by decomposing $\TT$ using basis operations and applying it after the initialisation, we can prepare the initial state without error. Actually, given an initial state that is close to $\ket{0}$, we can generate a complete set of linearly independent vectors $\{ \rket{\rho_k} \}$ using basis operations. With these vectors, we can decompose the initial state  without error as $\rket{\rho^{(0)}} = \sum_k q_k \rket{\rho_k}$.

A similar approach yields the corresponding result for measurement: For an observable $\rbra{Q}$ there will be some $\rbra{Q^{(0)}} = \rbra{Q} \TT$ where $\rbra{Q^{(0)}}$ is the error-free quantity. If an observable is close to $\sigma^{\rm z}$ then a linearly independent set $\{ \rbra{Q_j} \}$ can be generated; then the error-free observable $\rbra{Q^{(0)}} = \sum_j q_j \rbra{Q_j}$.

Circuits for QEM are shown in Fig.~\ref{fig:circuit}(a). Given quasi-probabilities, we can compute the corresponding probability in sampling circuits as shown in Sec.~\ref{sec:per-operation}. More details of QEM using basis operations are given in Appendix~\ref{app:decom}.

Using the same technique, we can also {\it increase} the error in an operation, as required by the alternative error extrapolation method for QEM. Instead of decomposing the error-free operation $\OO^{(0)}$ using $\OO$ and basis operations, we can also decompose the error-boosted operation $\OO_{\rm b}(r) = (1-r)\OO^{(0)} +r\OO$ ($r>1$) using $\OO$ and basis operations. It is similar for initial states and observables. We have noted that in the decomposition of an error-free operation, there are always some negative quasi-probabilities, i.e.~the $C$ factor is greater than $1$, which leads to greater time costs. But fortunately when we merely wish to decompose an {\it error-boosted} operation we can do so without introducing negative quasi-probability, e.g.~by boosting Pauli errors using Pauli gates~\cite{Li2017}.

\section{Quantum gate set tomography}

We can measure a set of initial states $\{ \rket{\bar{\rho}_k} \}$, observables $\{ \rbra{\bar{Q}_j} \}$ and operations $\{ \bar{\OO}_i \}$ (including basis operations) using GST~\cite{Merkel2013, Greenbaum2015}. These vectors and matrices with the bar notation describe the actual physical system. Because there are errors in both initial states and observables, and initialisation and measurement errors cannot be distinguished, we may not obtain exactly these vectors and matrices describing the actual physical system. Instead, the vectors and matrices obtained using GST are $\{ \rket{\hat{\rho}_k} \}$, $\{ \rbra{\hat{Q}_j} \}$ and $\{ \hat{\OO}_i \}$, which are estimations of $\{ \rket{\bar{\rho}_k} \}$, $\{ \rbra{\bar{Q}_j} \}$ and $\{ \bar{\OO}_i \}$, respectively.

If we know $\{ \rket{\bar{\rho}_k} \}$, $\{ \rbra{\bar{Q}_j} \}$ and $\{ \bar{\OO}_i \}$ because the physical system is well understood, we can directly use them in QEM. If our knowledge about the physical system is not enough, we can use GST to obtain $\{ \rket{\hat{\rho}_k} \}$, $\{ \rbra{\hat{Q}_j} \}$ and $\{ \hat{\OO}_i \}$. We will show that, although the estimations may not be exact, we can exactly correct errors by using these estimations in QEM.

Using the protocol in Refs.~\cite{Merkel2013, Greenbaum2015} (also see Appendix~{\ref{app:GST}}), the estimation of an operation and the actual physical operation are similar matrices, i.e.~$\hat{\OO}_i = T\bar{M}^{{\rm in} -1} \bar{\OO}_i \bar{M}^{\rm in} T^{-1}$, where $\bar{M}^{\rm in}$ is a matrix determined by initial states (i.e.~$\bar{M}^{\rm in}_{\sigma, k} = \rbraket{\sigma}{\bar{\rho}_k}$), and $T$ is an arbitrary invertible matrix. We note that $T$ and $\bar{M}^{\rm in}$ are independent of the operation $\bar{\OO}_i$, and $\bar{M}^{\rm in}$ cannot be determined by GST. By choosing $T$, we can obtain different estimations of the operation set. Similarly, $\rket{\hat{\rho}_k} = T\bar{M}^{{\rm in} -1} \rket{\bar{\rho}_k}$ and $\rbra{\hat{Q}_j} = \rbra{\bar{Q}_j} \bar{M}^{\rm in} T^{-1}$.

All operations are transformed by the same similarity transformation, and initial states and observables are also transformed accordingly. As a result, these estimations obtained by GST can exactly predict the expected value of an observable, i.e.~$\rbra{\bar{Q}_j} \bar{\OO}_N \cdots \bar{\OO}_1 \rket{\bar{\rho}_k} = \rbra{\hat{Q}_j} \hat{\OO}_N \cdots \hat{\OO}_1 \rket{\hat{\rho}_k}$. Therefore, we can directly use these estimations in QEM, and the similarity transformation does not lead to any computing error.

Using GST estimations in QEM, the actual operations realised in this way differ from operations without error, but the computing result is correct. To correctly obtain $\rbra{Q^{(0)}} \OO^{(0)} \rket{\rho^{(0)}}$, we decompose the initial state, the observable and the operation using $\{ \rket{\hat{\rho}_k} \}$, $\{ \rbra{\hat{Q}_j} \}$ and $\{ \hat{\OO}_i \}$ respectively. Here, $\rket{\rho^{(0)}}$, $\rbra{Q^{(0)}}$, $\OO^{(0)}$ and GST estimations are all known to us. The decompositions are $\rket{\rho^{(0)}} = \sum_k q_k \rket{\hat{\rho}_k}$, $\rbra{Q^{(0)}} = \sum_j q_j \rbra{\hat{Q}_j}$ and $\OO^{(0)} = \sum_i q_i \hat{\OO}_i$. Accordingly, we actually realise $\rket{\bar{\rho}^{(0)}} = \sum_k q_k \rket{\bar{\rho}_k}$, $\rbra{\bar{Q}^{(0)}} = \sum_j q_j \rbra{\bar{Q}_j}$ and $\bar{\OO}^{(0)} = \sum_i q_i \bar{\OO}_i$ in the physical system. We have $\rket{\bar{\rho}^{(0)}} = \bar{M}^{\rm in}T^{-1} \rket{\rho^{(0)}} $, $\rbra{\bar{Q}^{(0)}} = \rbra{Q^{(0)}} T\bar{M}^{{\rm in} -1}$ and $\bar{\OO}^{(0)} = \bar{M}^{\rm in} T^{-1} \OO^{(0)} T\bar{M}^{{\rm in} -1}$. Therefore, the physical system gives the computing result $\rbra{\bar{Q}^{(0)}} \bar{\OO}^{(0)} \rket{\bar{\rho}^{(0)}} = \rbra{Q^{(0)}} \OO^{(0)} \rket{\rho^{(0)}}$, i.e.~the desired error-free output. The cost of this adaption lies in the potential increase to the number of samples required, as shown in Fig.~\ref{fig:cost} and discussed in the caption.

We would like to remark that, when errors in actual operations are small, errors in estimations of operations are also small. If we take a proper strategy for choosing $T$, and errors in initial states and observables are small, the estimation of an operation $\hat{\OO}$ is close to the operation without error $\OO^{(0)}$ when the actual operation $\bar{\OO}$ is close to $\OO^{(0)}$. It is similar for estimations of initial states and observables. See Appendix~\ref{app:stability} for details.

\section{Estimation of the cost}

\begin{figure*}[tbp]
\centering
\includegraphics[width=1\linewidth]{\figpath 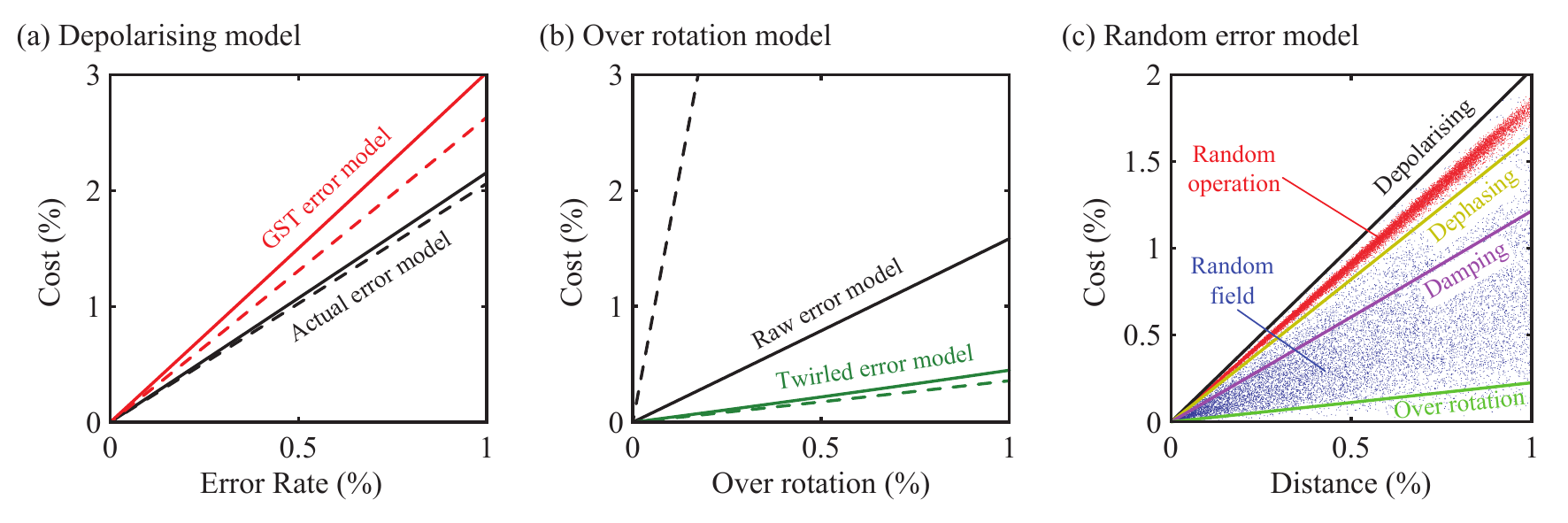}
\caption{
Cost ($C-1$) for correcting errors. We consider a universal set of operations, including the initialisation, measurement, single-qubit Clifford gates, a single-qubit non-Clifford gate, and a two-qubit entangling Clifford gate. Sixteen basis operations of each qubit can be generated using these operations. Every operation in the set has error, and the memory error is also included. We assume that qualities of the initialisation and single-qubit gates are $10$ times better than the measurement and two-qubit gates; also that the quality of the memory operation is $100$ times better. Details of the model are given in Appendix~\ref{app:model}. The cost for correcting error in each operation in the universal set is calculated, and the maximum cost over all operations is plotted in the figure. (a) For the depolarising error model, the cost is lower if we directly use actual operations to correct errors, and the cost is higher if we use gate-set-tomography (GST) estimations to correct errors. (b) For the over-rotation model, the cost is higher without using the Pauli twirling, and the cost is lower when the Pauli twirling is used. In (a) and (b), solid curves correspond to the compensation method (with the optimised $\lambda$), and dashed curves correspond to the inverse method. (c) The cost as a function of the distance between operations with errors and operations without error. For the operation with error $\OO$ and the operation without error $\OO^{(0)}$, the distance is $\epsilon_\OO = \norm{\OO-\OO^{(0)}}_{\rm max}$. The x-axis illustrates the maximum distance over all operations in the universal set. In (b) and (c), we always use GST estimations. In (c), Pauli twirling and the inverse method are used for all the data. Pauli twirling is applied to the measurement and two-qubit gate, and the inverse method is only applied to the two-qubit gate, while errors in other operations are corrected using the compensation method. We remark that usually the maximum distance and the maximum cost are given by the two-qubit gate.
}
\label{fig:cost}
\end{figure*}

{In general, when the error in an operation is more significant, there is a higher cost for mitigating the error (to a given level of suppression)}. We take $\epsilon_\OO = \norm{\OO - \OO^{(0)}}_{\rm max}$ as the measure of the error severity in the operation, where $\OO$ ($\OO^{(0)}$) is the $n$-qubit operation with (without) error. An upper bound of the cost for correcting error in $\OO$ is
\begin{eqnarray}
C_\OO - 1 \leq \frac{16^{2n} \epsilon_\OO}{[s_{\rm min}(A^{(0)}) - 16\epsilon_{\rm max}]^n}.
\end{eqnarray}
Here, $\epsilon_{\rm max}$ is the maximum error in all basis operations for all $n$ qubits, and $s_{\rm min}(A^{(0)}) = \frac{1}{2}(13-3\sqrt{17}) \approx 0.315$. Similar upper bounds can be obtained for correcting errors in initial states and observables. See Appendix~\ref{app:upper} for details.

There are several ways for reducing the cost. The upper bound of the cost is obtained using the compensation method and taking $\lambda = 1$. In general, we can optimise the value of $\lambda$ or use the inverse method to minimise the cost. For example, for the depolarising error model (see Appendix~\ref{app:model}), the cost of using the inverse method is lower than using the compensation method [see Fig.~\ref{fig:cost}(a)]. We remark that, to obtain data for the compensation method in Fig.~\ref{fig:cost}, we have optimised the value of $\lambda$. If we use estimations obtained from GST to correct errors, we can optimise the $T$ matrix to minimise the cost. In Fig.~\ref{fig:cost}(a), we can find that, without optimising $T$ matrices, the cost using estimations obtained from GST is higher than using actual operations. If we choose the matrix in the form $T = \otimes_{m=1}^n T_m$, where $T_m$ is a $4$-dimensional real matrix corresponding to the $m^\text{th}$ qubit, there are total $16n$ parameters to be optimised for a $n$-qubit quantum computer, which is a non-trivial task. Under some reasonable conditions, we can also use the Pauli twirling~\cite{Knill2004, Wallman2015, OGorman2016} to reduce the cost.

\subsection*{Pauli twirling}

In many quantum computing systems, e.g.~superconducting qubits~\cite{Barends2014} and ion traps~\cite{Harty2014, Lucas, Wineland}, the fidelity of single-qubit gates is much better than the fidelity of two-qubit gates, and usually a state can be initialised with a high fidelity while the fidelity of measurement is worse. In this section, we focus on the case that error rates of initialisation and single-qubit gates are much lower than error rates of two-qubit gates and measurement.

If the error rate of initialisation is low (much lower than the error rate of measurement), we know how to choose $T$ so that the estimation of an operation obtained from GST is close to the actual operation. We cannot exactly estimate operations using GST, because we cannot distinguish initialisation and measurement errors. If we treat all errors in the initialisation and measurement as measurement error (which corresponds to $T = M^{{\rm in}(0)\otimes n}$ in Appendix~\ref{app:stability}), the difference between the estimation and the actual operation is only determined by the initialisation error. Therefore, if the initialisation is high-fidelity, the estimation obtained in this way and the actual operation are close.

Because the set of basis operations includes Pauli gates, it is easy to use basis operations to correct Pauli errors. By using the Pauli twirling, we can convert the error in a two-qubit entangling Clifford gate to Pauli error~\cite{Knill2004, Wallman2015, OGorman2016}, which is achieved by applying Pauli gates before and after the two-qubit gate. This treatment of the error is feasible only if the fidelity of Pauli gates is much better than the two-qubit gate, otherwise Pauli gates cause significant new errors, which may not be Pauli error, on the two-qubit gate. In Fig.~\ref{fig:cost}(b), we can find that the cost can be significantly reduce by using the Pauli twirling for the over-rotation error model (see Appendix~\ref{app:model}).

In Fig.~\ref{fig:cost}(c), costs of different error models are compared, including the depolarising model, pure-dephasing model, amplitude-damping model and the over-rotation model. We also randomly generated many other error models, please see Appendix~\ref{app:model} for details of these error models. For a random-operation model, we randomly generate an operation close to the ideal error-free operation, and we find that the cost is approximately the cost of the depolarising model. For a random-field model, we randomly generate a Hamiltonian that drives the erroneous evolution, and the cost is between the depolarising model and over-rotation model.

\RED{From Fig.~\ref{fig:cost}(c) we see that the cost of quantum error mitigation varies according to the error model but is generally upper-bounded by the case of depolarising noise, over the range of noise levels shown here. (Note that other models can exceed the cost of the the depolarising model if we use even lower fidelity gates). For the depolarising model, the cost for mitigating error in a two-qubit entangling gate is $C-1 \simeq a\epsilon$, where $\epsilon$ is the error rate and the factor $a$ is between $2$ and $3$ [see Fig.~\ref{fig:cost}(a)]. If errors in initialisation and single-qubit gates are negligible, or if the matrix $T$ is optimised to minimise the cost, the factor $a$ can approach $2$. Accepting the depolarising model as an approximate upper bound, we can estimate the overall cost in a quantum algorithm. Suppose the total number of gates in a quantum algorithm is $N$, the overall amplification of the standard deviation (uncertainty of the computing result) is $(1+2\epsilon)^N$. Therefore, $(1+2\epsilon)^{2N}$ times more repetitions of the experiment are required in order to reduce the standard deviation. We are interested in the case that $N$ is large but $\epsilon$ is small, therefore, $(1+2\epsilon)^{2N} \sim e^{4N\epsilon}$. As a rule of thumb we might take $N\epsilon = 2$ as a limit for acceptable scenarios, since then $e^{4N\epsilon} \approx 3,000$. However, larger overhead factors may be acceptable depending on the speed of the quantum computer.}

\section{Numerical simulation}

\begin{figure}[tbp]
\centering
\includegraphics[width=1\linewidth]{\figpath 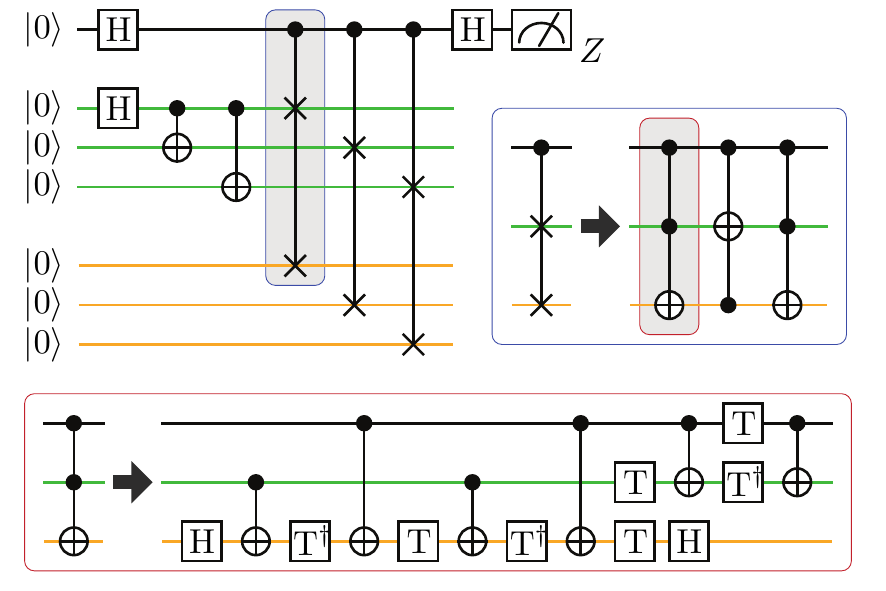}
\caption{
Swap-test circuit. The first qubit (denoted black) is a probe qubit, and the expected value of $Z$ gives the overlap between states of two qubit groups (denoted green and orange, respectively). Green qubits are prepared in the GHZ state $(\ket{00\cdots}+\ket{11\cdots})/\sqrt{2}$, and orange qubits are prepared in $\ket{00\cdots}$. Therefore, the ideal expected value of $Z$ is $0.5$.
}
\label{fig:swap}
\end{figure}

\RED{In our numerical simulation, we apply QEM to the {\small SWAP}-test circuit~\cite{Ekert2002} shown in Fig.~\ref{fig:swap}, in which we realise each controlled-{\small SWAP} gate using Toffoli gates and realise each Toffoli gate using $T$ gates, $T^\dag$ gates, Hadamard gates and controlled-{\small NOT} gates~\cite{Nielsen2010}. We note with interest that very recently, the implementation of a {\small SWAP} test using shallow circuit has been proposed~\cite{shallowswap}. However, for present purposes it is not essential to use an optimised realisation of the SWAP circuit; its role is simply to act as a real test case for our technique and indeed the considerable depth of our non-optimal circuit is helpful here. The number of gates scales as $23 N_{\rm q} - 21$, where $N_{\rm q}$ is the number of qubits (e.g.~$N_{\rm q}= 7$ in Fig.~\ref{fig:swap}). Without error, the expected value of the observable $Z$ ($\sigma^{\rm z}$ of the probe qubit) in the {\small SWAP}-test circuit in Fig.~\ref{fig:swap} is $0.5$.}

\begin{figure*}[tbp]
\centering
\includegraphics[width=0.96\linewidth]{\figpath 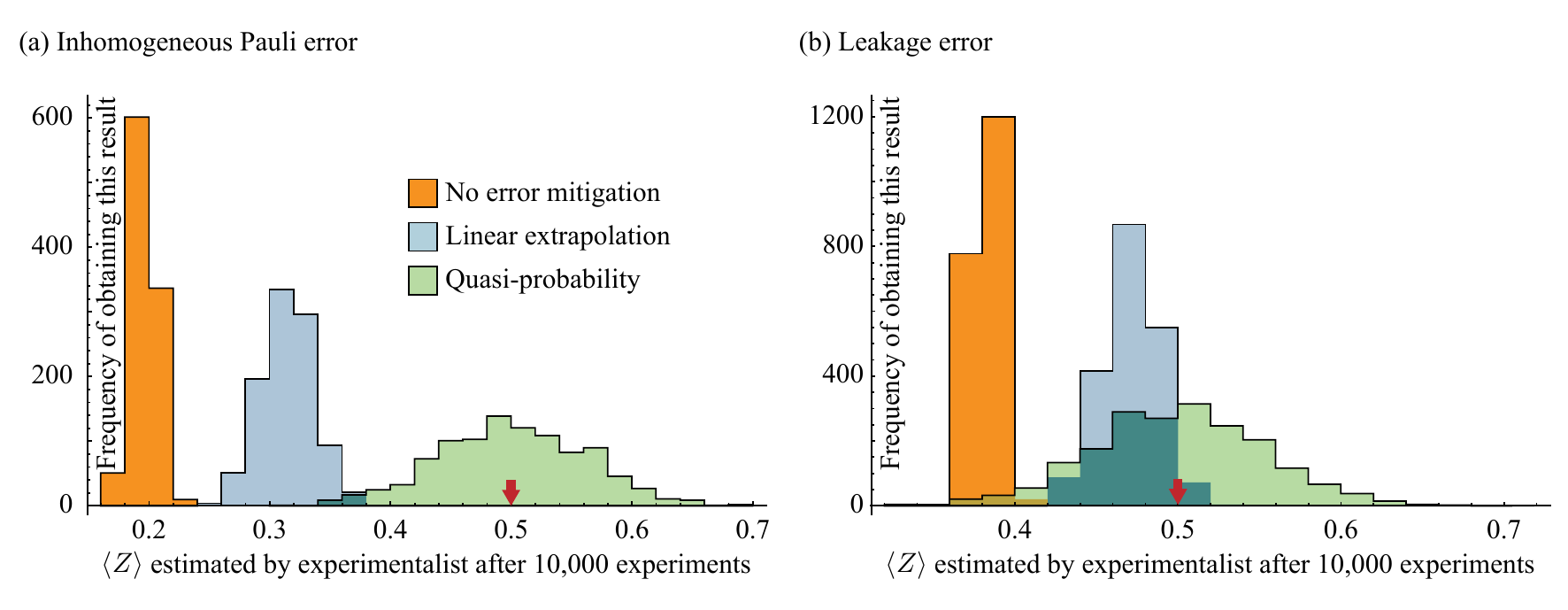}
\caption{
Histograms of the estimation of $\mean{Z}$ using quantum computers with inhomogeneous Pauli error and leakage error. For the inhomogeneous Pauli error model, the {\small SWAP}-test circuit involving $19$ qubits is simulated: one qubit is the probe qubit, and each group has $9$ qubits. For the leakage error model, the {\small SWAP}-test circuit involving fewer qubits ($15$ qubits) is simulated, because in the numerical simulation we need to use an additional qubit to introduce the leakage process. The ideal value of $\mean{Z}$ is marked by the red arrow.
}
\label{fig:Data}
\end{figure*}

We consider error models according to which the same noise $\mathcal{E}$ is applied after the initialisation to the state $\ket{0}$, before the measurement, and before and after each gate. For the controlled-{\small NOT} gate, the noise applied is $\mathcal{E} \otimes \mathcal{E}$ on two qubits. We remark that basis operations are also affected by noise likewise. We consider two types of noise: inhomogeneous Pauli error and leakage error, which can be respectively described as
$$
\mathcal{E}_{\rm inh} = (1-p_{\rm x}-p_{\rm y}-p_{\rm z})[\openone] + p_{\rm x}[\sigma^{\rm x}] + p_{\rm y}[\sigma^{\rm y}] + p_{\rm z}[\sigma^{\rm z}]
$$
and
$$
\mathcal{E}_{\rm leak} = [\ketbra{0}{0} + \sqrt{1-p}\ketbra{1}{1}],
$$
where $p_{\alpha}$ is the probability of the error $[\sigma^{\alpha}]$, and $p$ is the probability of the leakage error from the state $\ket{1}$. It is worth mentioning that the leakage error is a non-trace-preserving error. In our simulations, we set $p_{\rm x} = p_{\rm y} = 0.0001$, $p_{\rm z} = 0.0006$, and $p = 0.0008$. Thus in both models the total error rate is $0.08\%$ for initialisation and measurement, $0.16\%$ for single-qubit gates and $0.32\%$ for two-qubit gates, which is achievable with two-qubit gates in ion traps~\cite{Lucas} and can be far surpassed for one-qubit gates~\cite{Harty2014}. Moreover, with these numbers the expected total number of error events in circuits of the depth and breadth that we consider here is approximately unity; this is a challenging domain for error mitigation.

In addition to quasi-probability decomposition \RED{(see Appendix~\ref{app:instruction} for an instruction of the implementation)}, we also study the extrapolation technique introduced in Ref.~\cite{Li2017}. The expected value of $Z$ obtained by running the {\small SWAP}-test circuit in a quantum computer with noise depends on the error rate, i.e.~it is a function that can be denoted as $\mean{Z}(\epsilon)$, where $\epsilon$ is the overall error rate. For our first set of numerical experiments we consider {\it linear extrapolation} to the error-free value $\mean{Z}(0)$ as follows: We obtain the expected value $\mean{Z}(\epsilon_0)$ with the lowest attainable error rate $\epsilon_0$, and by increasing error rate to $r\epsilon_0$ with $r>1$, we obtain another expected value $\mean{Z}(r\epsilon_0)$. Using these two values, we can infer $\mean{Z}(0) = (r\mean{Z}(\epsilon_0) - \mean{Z}(r\epsilon_0))/(r-1)$ as shown in Fig.~\ref{fig:circuit}(b), which is the final estimation of $\mean{Z}$. Here, we set $r = 2$.

The first set of numerical results are shown in Fig.~\ref{fig:Data}. We assume that the experimentalist makes her overall estimate of the $\mean{Z}$ after she performs $10^4$ individual experiments. {We take this number of runs as a fixed constraint (effectively, we are constraining her overall time resource), and she may choose to} employ those runs using one of three alternative approaches: no error correction, linear extrapolation, and quasi-probability decomposition (using basis operations and incorporating GST). In each experiment the {\small SWAP}-test circuit or its variant for the purpose of QEM is implemented. Because of the finite number of samples, the estimation is stochastic. Therefore, in our numerical simulation we perform the appropriate series of $10^4$ experiments, mirroring the actions of the experimentalist, and then we repeat $\geq 1,000$ times in order to determine the distribution of final estimations that may be obtained. The distribution for each case is plotted in Fig.~\ref{fig:Data}.

\begin{figure*}[tbp]
\centering
\includegraphics[width=0.96\linewidth]{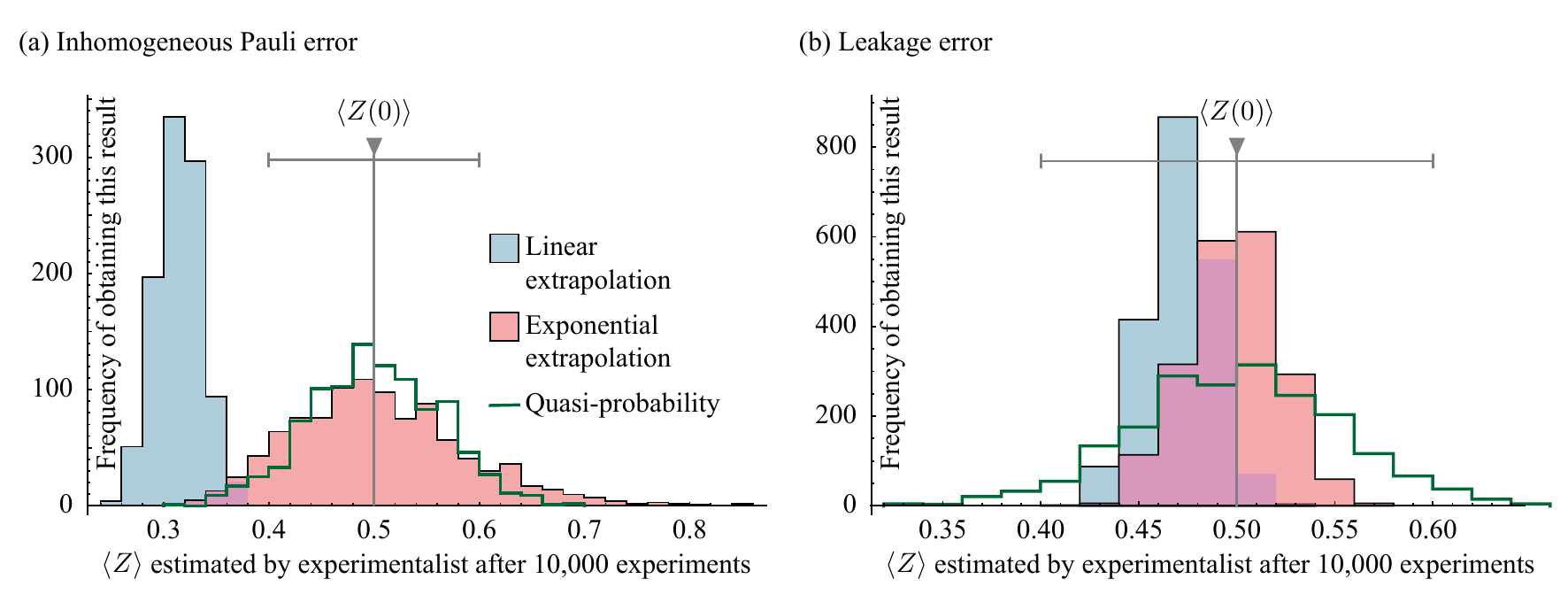} 
\caption{
Comparison of optimised quantum error mitigation techniques: The green outlines correspond to the quasi-probability technique while solid histograms correspond to the extrapolation technique using a presumption of an underlying linear (blue) or exponential (red) dependence. \RED{For the inhomogeneous Pauli error model, the {\small SWAP}-test circuit involving $19$ qubits is simulated. For the leakage error model, the {\small SWAP}-test circuit involving fewer qubits ($15$ qubits) is simulated.} The horizontal axis is the estimate of $\mean{Z}$ that an experimentalist who performs $10^4$ experiments will obtain. Ideally the circuit produces $\mean{Z} = 0.5$. (a) The left panel corresponds to physical errors of the inhomogeneous Pauli type, while the right panel (b) corresponds to physical leakage errors. Note that the horizontal scale differs between the two panels; a grey bar showing the scale from $0.4$ to $0.6$ appears in both figures to facilitate comparison. For either type of noise, it is clear that exponential extrapolation mitigates noise more than linear extrapolation.
}
\label{fig:bothOpt}
\end{figure*}

We can observe that both QEM approaches can improve the result, i.e.~the corresponding distributions are shifted closer to the ideal value $0.5$ compared to the approach without QEM. For the inhomogeneous Pauli error model, the means of distributions are at $0.1961$, $0.3415$, and $0.5011$ for the three approaches, respectively. The distribution of the quasi-probability approach is centered at the ideal value, which clearly shows its {desirable property of completely removing any systematic bias}. However, the distribution is wider (as we expected) compared to other two approaches. A more fair metric would be the expected absolute error versus ideal value (i.e.~$\overline{\abs{\mean{Z}-0.5}}$). {Given an ideal error-free computer and $10^4$ trials, this metric would evaluate to $0.006910$}. Using the error-prone computer with our three protocols the three corresponding values are $0.3039$, $0.1853$ and $0.0491$. Similarly, for the leakage error model, the means for three approaches now lie at $0.3819$, $0.4710$, $0.5007$, while the expected absolute error evaluates to  $0.1181$, $0.0294$, and $0.0434$.

From these results it may appear that (given a large but reasonable number of samples) the quasi-probability technique outperforms the extrapolation method, with the latter unable to approach the mean of the error-free circuit. However, here the extrapolation method was limited to linear interpolation whereas the physical error rates are high enough that the linear assumption is poor. One could fit a higher order polynomial using more data points (here, we have only used two: one derived from the actual lowest possible error rate and one boosted to twice the error rate); however since we are limiting the total number of experimental runs to $10^4$ this would lead to greater noise in each data point. Moreover, as we now argue, the underlying tend is likely to be well-approximated by an exponential decay rather than a polynomial one (i.e.~the expected value of the observable falls exponentially with the physical error rate) and two data points will suffice to estimate the zero-error observable under that assumption.

In Fig.~\ref{fig:bothOpt} we show the results  when the experimentalist indeed assumes that the expected value $\mean{Z}(\epsilon)$ changes exponentially with respect to the error rate $\epsilon$ and converges to $0$ in the limit of $\epsilon \rightarrow \infty$. Then she will infer the error-free value as
$$
\mean{Z}(0) = \mean{Z}(\epsilon_0)^{\frac{r}{r-1}} \mean{Z}(r\epsilon_0)^{\frac{1}{1-r}}
$$
Here we take $r=2$.

As shown in Fig.~\ref{fig:bothOpt}, the distribution of the final result using the exponential extrapolation approaches the ideal value of $\mean{Z}$ (which is $0.5$ for the {\small SWAP}-test circuit) much better than the linear extrapolation. Given the same $10^4$ experimental runs, the mean of the experimentalist's estimate is now $0.5111$ for the inhomogeneous Pauli error model and $0.4986$ for the leakage error model. These numbers almost rival those of quasi-probability technique but do so with a smaller variance. The expected absolute error for inhomogeneous Pauli error and leakage error are 0.06501 and 0.01882, respectively. For the latter, the expected absolute error comes within a factor of three of the shot-noise limit that would be achieved by error-free ideal hardware ($0.00691$). This is despite the fact that our error-burdened circuits have error rates corresponding to at least one error event per circuit. We emphasise that this suppression results purely from the QEM protocol i.e. it is achieved at no cost in terms of the qubit count or the total number of runs (constrained to $10^4$).

\begin{figure*}[tbp]
\centering
\includegraphics[width=0.96\linewidth]{\figpath 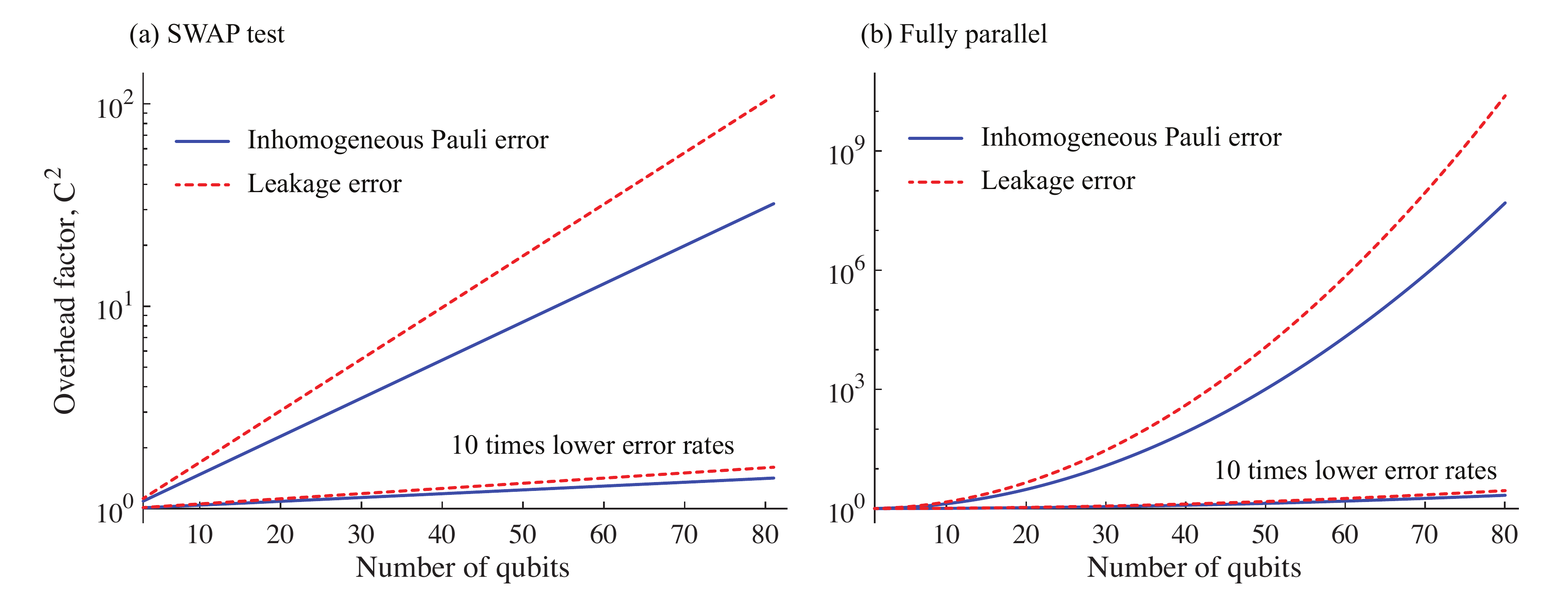}
\caption{
\RED{The graphs show the cost of matching the performance of an ideal noiseless circuit with a noisy circuit, using the quasi-probability method. The vertical axis ($C^2$) is a multiplicative factor indicating how many more repetitions of the circuit execution are requited. In each graph the upper pair of lines correspond to error rates achievable in ion trap experiments~\cite{Harty2014, Lucas}, i.e.~the error rate of two-qubit gate is $0.1\%$ and error rates of single-qubit operations are $0.01\%$. The lower pair of lines indicate the result of reducing these error rates by a factor of ten. The left panel corresponds to the SWAP-test circuit. The right panel corresponds to a circuit where every qubit is actively gated in every time step, and the number of steps equals to the number of qubits.}
}
\label{fig:SwapTestCost.pdf}
\end{figure*}

\RED{Due to the limited power of classical computer we utilised, our exact numerical simulations did not involve more than 19 qubits. However, it is of course very interesting to assess the relevance of our techniques to quantum computing using over 50 qubits, which is in the so called `quantum supremacy' regime. Therefore, we estimate the cost of quantum error mitigation in the {\small SWAP}-test circuit, using the same error models in our numerical simulation and error rates achievable in ion trap experiments~\cite{Harty2014, Lucas}, i.e.~the error rate of two-qubit gate is $0.1\%$ and error rates of single-qubit operations are $0.01\%$. Take for example the {\small SWAP} test with $N_{\rm q} = 51$ qubits (the number of gates is $1,152$). For the inhomogeneous Pauli error model, the overall cost is $C = 2.956$, which implies that we can attain the same computing precision as the ideal case if we have $C^2 = 8.738$ times more repetitions of the experiment, which is experimentally feasible. For the leakage error model, the cost for the $51$-qubit {\small SWAP} test is $C = 4.338$, which means $C^2 = 18.818$ times more repetitions. A plot showing how the cost scales versus qubit count is shown in Fig.~\ref{fig:SwapTestCost.pdf}.

We also evaluate $C^2$ for a fully paralleled circuit, whose circuit depth is $N_{\rm q}$, and each layer has $N_{\rm q}/2$ single qubit gates and $N_{\rm q}/4$ controlled-NOT gates, which means the quantum circuit has $N_{\rm q}^2/2$ single qubit gates and  $N_{\rm q}^2/4$ controlled-NOT gates. As single qubit gates, we use $T$ gate, $S$ gate and Hadamard gate, because these gates plus controlled-NOT gate constitute a universal gate set, and we equally assign the number of qubits to these three types of single qubit gates. We plot $C^2$ versus the number of qubits for the {\small SWAP}-test circuit and the fully paralleled circuit in Fig.~\ref{fig:SwapTestCost.pdf}. We observe that for the {\small SWAP} test circuit, it is feasible to venture into the `supremacy' regime with today's best fidelities; for the more demanding case of full parallelism (so that the gate count scales as $N_{\rm q}^2$) we see that today's error rates would not suffice much beyond $50$ qubits, but that error rates ten times lower would easily suffice for $80$ qubits and beyond.}

\section{Intuition for Exponential extrapolation}
\label{EEsection}

Intuitively, the explanation for the success of the exponential extrapolation is as follows. We express the $i^\text{th}$ noise event occurring in the quantum circuit as
\begin{eqnarray}
\mathcal{E}_i(\epsilon) = (1-\epsilon) [I] + \epsilon \mathcal{E}'(\epsilon)_i,
\end{eqnarray}
where $\mathcal{E}'(\epsilon)$ is the error component. The only assumption is that the error component only weakly depends the error rate $\epsilon$ (see Appendix~\ref{app:errorCom}). Now, for simplification we ignore the computing operations, which do not affect our general argument. The total noise that the entire quantum circuit experiences is
\begin{eqnarray}
\prod_{i=1}^N \mathcal{E}_i = \prod_{i=1}^N [(1-\epsilon) [I] + \epsilon \mathcal{E}'_i].
\end{eqnarray}
Here, $N$ is the total number of the noise-burdened operations. Expanding the overall noise, we get
\begin{eqnarray}
\prod_{i=1}^N \mathcal{E}_i = \sum_{n=1}^N \binom{N}{n} (1-\epsilon)^{N-n}\epsilon ^n \mathcal{X}_n,
\end{eqnarray}
where 
\begin{eqnarray}
\mathcal{X}_n = {\binom{N}{n}}^{-1} \times \left(
\begin{array}{c}
\text{the sum of terms where} \\
\mathcal{E}' \text{ appears for } n \text{ times}
\end{array}
\right).
\end{eqnarray}
Note that the coefficient of $\mathcal{X}_n$ in the overall noise corresponds to a binomial distribution, which can be approximated by the Poisson distribution. We have 
\begin{eqnarray}
\prod_{i=1}^N \mathcal{E}_i = e^{-N\epsilon} \sum_{n=0}^N \frac{(N\epsilon)^n}{n!} \mathcal{X}_n,
\end{eqnarray}
We can find that the impact of the overall noise on the expected value of some observable is proportional to $e^{-N\epsilon}$, which implies that exponential extrapolation works better than linear extrapolation.

\section*{Conclusions}

We have demonstrated that, following our protocol step by step, an experimentalist can derive an algorithm to run on a noisy quantum computer so as to estimate an output observable with zero bias versus the ideal observable. The experimentalist does not require any prior knowledge of the physical property of the noise, and the only condition is that the noise is localised and Markovian. For this purpose, we have shown that quantum gate set tomography is a perfect tool for measuring the noise in a quantum computer, if the aim is only to compensate the effect of the noise in quantum computing; and we also have shown that single-qubit Clifford gates and measurement can derive a complete set of operations that can compensate any noise in quantum computing. 

The price of using such a systematic method to negate computing errors is that the quantum computation needs to run for a longer time than an error-free system. We verify the protocol with numerical simulations of up to 19 qubits, in which an alternative method, i.e.~exponential error extrapolation, is introduced and studied. We find that the estimation using exponential error extrapolation is also very accurate, while the computing time could be shorter. An approach combining two methods may optimise both accuracy and efficiency. 

\RED{In Appendix~\ref{app:instruction} we describe in detail the steps that an experimentalist would take in order to realise the quasi-probability method. We hope that this compact summary, presented in a single section, will indeed be useful to researchers who are interested in demonstrating the QEM technique with their hardware. 
}

Our general conclusion is that these quantum error mitigation techniques can dramatically enhance the performance of quantum computers, especially at the small-to-medium scale where full code-based quantum error correction is impossible. Our simulations have considered circuits up to $19$ qubits, but with error rates considerably {\it worse} than the state of the art. Extrapolating from the trends that we observe in these smaller systems, we anticipate that hybrid algorithms involving $50+$ qubits, i.e. beyond the reach of classical emulation, will benefit from QEM techniques if the hardware fidelity matches today's state-of-the-art error or modestly improves upon it.

\begin{acknowledgments}
This work was supported by the EPSRC National Quantum Technology Hub in Networked Quantum Information Technologies. SE is supported by Japan Student Services Organization (JASSO) Student Exchange Support Program (Graduate Scholarship for Degree Seeking Students). YL is also supported by NSAF (Grant No. U1730449). The authors would like to acknowledge the use of the University of Oxford Advanced Research Computing (ARC) facility in carrying out this work. http://dx.doi.org/10.5281/zenodo.22558.
\end{acknowledgments}

\appendix

\section{Pauli transfer matrix}
\label{app:PTM}

A state $\rho$ can be expressed as a real column vector
\begin{eqnarray}
\rket{\rho} = \left[
\begin{array}{ccc}
\cdots & \rho_\sigma & \cdots
\end{array}
\right]^{\rm T},
\end{eqnarray}
where the vector element is
\begin{eqnarray}
\rho_\sigma = \Tr(\sigma \rho),
\end{eqnarray}
$\sigma \in \{ \openone, \sigma^{\rm x}, \sigma^{\rm y}, \sigma^{\rm z} \}^{\otimes n}$ is a Pauli operator, and $d = 2^n$ is the dimension of the Hilbert space. Similarly, an observable (i.e.~Hermitian operator) $Q$ can be expressed as a real row vector
\begin{eqnarray}
\rbra{Q} = \left[
\begin{array}{ccc}
\cdots & Q_\sigma & \cdots
\end{array}
\right],
\end{eqnarray}
where the vector element is
\begin{eqnarray}
Q_\sigma = d^{-1} \Tr(\sigma Q).
\end{eqnarray}
Here, we use notations $\rbra{\cdot}$ and $\rket{\cdot}$ to denote real row and column vectors, respectively. A physical operation $\OO$ (i.e.~$\OO(\rho) = \sum_k E_k \rho E_k^\dag$) can be expressed as a real square matrix
\begin{eqnarray}
\OO_{\sigma, \tau} = d^{-1} \Tr[\sigma \OO(\tau)],
\end{eqnarray}
where $\sigma, \tau \in \{ \openone, \sigma^{\rm x}, \sigma^{\rm y}, \sigma^{\rm z} \}^{\otimes n}$ are Pauli operators. If $\rho' = \OO (\rho)$, we have $\rket{\rho'} = \OO \rket{\rho}$.

\section{Decomposition of controlled-{\small NOT} gate using basis operations}
\label{app:CNOT}

\RED{The controlled-{\small NOT} gate reads
\begin{eqnarray}
\Lambda_{\rm X} = \frac{\openone + \sigma^{\rm z}}{2} \otimes\openone + \frac{\openone - \sigma^{\rm z}}{2} \otimes\sigma^{\rm x}.
\end{eqnarray}
The controlled-{\small NOT} gate can be decomposed as
\begin{eqnarray}
[\Lambda_{\rm X}] &=& \frac{1}{2}( [\openone\otimes\sigma^{\rm x}] + [\sigma^{\rm z}\otimes\openone] - [\openone\otimes R_{\rm x}] \notag \\
&& - [R_{\rm z}\otimes\openone] - [\sigma^{\rm z}\otimes R_{\rm x}] - [R_{\rm z}\otimes\sigma^{\rm x}]) \notag \\
&& + [\sigma^{\rm z}\otimes\sigma^{\rm x}] + [R_{\rm z}\otimes R_{\rm x}] + [\openone\otimes\pi_{\rm x}] \notag \\
&& + [\pi_{\rm z}\otimes\openone] - [\sigma^{\rm z}\otimes\pi_{\rm x}] - [\pi_{\rm z}\otimes\sigma^{\rm x}].
\end{eqnarray}
Then, the corresponding cost is given by $C = 9$.}

\section{Error threshold of basis operations}
\label{app:BOth}

For two real matrices $A^{(0)}$ and $A$ and a non-zero real vector $x$, we have
\begin{eqnarray}
\norm{A^{(0)}x}_2 = \sqrt{x^{\rm T} A^{(0) {\rm T}} A^{(0)} x} \geq s_{\rm min}(A^{(0)}) \norm{x}_2,
\end{eqnarray}
where $s_{\rm min}(A^{(0)})$ is the minimum singular value of $A^{(0)}$. We also have
\begin{eqnarray}
\norm{(A-A^{(0)})x}_2 \leq \norm{A-A^{(0)}}_2 \norm{x}_2.
\end{eqnarray}
Therefore,
\begin{eqnarray}
\norm{Ax}_2 &\geq & \norm{A^{(0)}x}_2 - \norm{(A-A^{(0)})x}_2 \notag \\
&\geq & (s_{\rm min}(A^{(0)}) - \norm{A-A^{(0)}}_2) \norm{x}_2.
\label{ineq_I}
\end{eqnarray}
If $\norm{A-A^{(0)}}_2 < s_{\rm min}(A^{(0)})$, $\norm{Ax}_2$ is always positive (non-zero), i.e.~$A$ is invertible.

Now, $A$ is the matrix formed by basis operations with error as defined in Eq.~(\ref{eq:A}), and $A^{(0)}$ is the matrix formed by basis operations without error. Because $\det(A^{(0)}) = 16$, $A^{(0)}$ is invertible, i.e.~basis operations without error are linearly independent. The minimum singular value is $s_{\rm min}(A^{(0)}) = \frac{1}{2}(13-3\sqrt{17})$. Because $\norm{A-A^{(0)}}_2 \leq 16 \norm{\tilde{A} - A}_{\rm max} = 16\epsilon_{\rm max}$, the matrix $A$ is invertible if $\epsilon_{\rm max} < \frac{1}{16}s_{\rm min}(A)$.

\section{Decomposition using basis operations}
\label{app:decom}

We consider the $n$-qubit operation $\EE$. For each qubit, there is a set of basis operations $\{ \BB_{m,i} \vert i=1,\ldots,16\}$, where $m=1,\ldots,n$ is the label of the qubit. For each set of basis operations, there is a matrix $A$ as defined in Eq.~(\ref{eq:A}). We use $A_m$ to denote the matrix of the $m^\text{th}$ qubit.

The operation $\EE$ is decomposed as
\begin{eqnarray}
\EE = \sum_{i_1 = 1}^{16} \cdots \sum_{i_n = 1}^{16} q_{i_1,\ldots,i_n} \BB_{1,i_1}\otimes\cdots\otimes \BB_{n,i_n}.
\label{eq:decomE}
\end{eqnarray}
Coefficients form a $16^n$-dimensional vector
\begin{eqnarray}
q = \left(
\begin{array}{c}
q_{1,1,\cdots ,1,1} \\
\vdots \\
q_{1,1,\cdots ,1,16} \\
\vdots \\
q_{16,16,\cdots ,16,1} \\
\vdots \\
q_{16,16,\cdots ,16,16}
\end{array}
\right),
\label{eq:q}
\end{eqnarray}
Therefore, the decomposition is given by $q = (A_1 \otimes\cdots\otimes A_n)^{-1} E$, where $E$ is a $16^n$-dimensional vector corresponding to $\EE$.

We choose the order of Pauli operators, i.e. the order of bases of Pauli transfer matrices $\{ \BB_{m,i} \vert i=1,\ldots,16\}$, as $\openone$, $\sigma^{\rm x}$, $\sigma^{\rm y}$ and $\sigma^{\rm z}$ (which are also denoted as $I$, $X$, $Y$ and $Z$, respectively). Then, to be consistent with $A_1 \otimes\cdots\otimes A_n$, we have
\begin{eqnarray}
E = \left(
\begin{array}{c}
\EE_{I_1I_2\cdots I_{n-1}I_n, I_1I_2\cdots I_{n-1}I_n} \\
\EE_{I_1I_2\cdots I_{n-1}I_n, I_1I_2\cdots I_{n-1}X_n} \\
\vdots \\
\EE_{I_1I_2\cdots I_{n-1}Z_n, I_1I_2\cdots I_{n-1}Y_n} \\
\EE_{I_1I_2\cdots I_{n-1}Z_n, I_1I_2\cdots I_{n-1}Z_n} \\
\vdots \\
\EE_{Z_1Z_2\cdots Z_{n-1}I_n, Z_1Z_2\cdots Z_{n-1}I_n} \\
\EE_{Z_1Z_2\cdots Z_{n-1}I_n, Z_1Z_2\cdots Z_{n-1}X_n} \\
\vdots \\
\EE_{Z_1Z_2\cdots Z_{n-1}Z_n, Z_1Z_2\cdots Z_{n-1}Y_n} \\
\EE_{Z_1Z_2\cdots Z_{n-1}Z_n, Z_1Z_2\cdots Z_{n-1}Z_n}
\end{array}
\right).
\label{eq:E}
\end{eqnarray}
Here, $\alpha_m$ ($\alpha = I,X,Y,Z$) is a Pauli operator of the $m^\text{th}$ qubit.

The state of a qubit is represented by a $4$-dimensional real vector. To decompose the initial state of a qubit without error $\rket{\rho^{(0)}}$, we need four linearly independent initial states. If the qubit can be initialised in the state $\ket{0}$, we can choose the set of four states as $\{ \rho_k^{(0)} \} = \{ \ket{0}, \ket{1}, \frac{1}{\sqrt{2}}(\ket{0}+\ket{1}), \frac{1}{\sqrt{2}}(\ket{0}+i\ket{1}) \}$. These four states can be obtained by applying basis-adjusting operations (Clifford gates) $\{ [\openone], [R_{\rm x}], [R_{\rm x}]^2, [R_{\rm z}][R_{\rm x}] \}$ on the initial state $\ket{0}$. Because of the error in the state $\ket{0}$ and errors in basis-adjusting operations, the prepared four states $\{ \rho_k \}$ are not exactly states $\{ \rho_k^{(0)} \}$. When the overall error is small, states $\{ \rho_k \}$ are still linearly independent. We introduce the matrix $M^{\rm in}_{\sigma, k} = \rbraket{\sigma}{\rho_k}$, and $M^{{\rm in} (0)}$ is the matrix corresponding to $\{ \rho_k^{(0)} \}$. States $\{ \rho_k \}$ are linearly independent if $M^{\rm in}$ is invertible. Similar to the analyse of the linear independence of basis operations (i.e.~the invertibility of the matrix $A$, see Appendix~\ref{app:BOth}), we have that $M^{\rm in}$ is always invertible if $\norm{ M^{\rm in} - M^{{\rm in} (0)} }_{\rm max} < \frac{1}{4}s_{\rm min}(M^{{\rm in} (0)}) = \frac{1}{8}\sqrt{\frac{5-\sqrt{17}}{2}} \simeq 0.0828$. The initial state without error is decomposed as $\rket{\rho^{(0)}} = \sum_{k=1}^4 q_k \rket{\rho_k}$. Coefficients form a $4$-dimensional column vector $q = [q_1 ~ q_2 ~ q_3 ~ q_4]^{\rm T}$. The decomposition is given by $q = M^{{\rm in} -1} \rket{\rho^{(0)}}$.

Similarly, an observable of a qubit is also represented by a $4$-dimensional real vector. To decompose the observable of a qubit without error $\rbra{Q^{(0)}}$, we need four linearly independent observables. If $\sigma^{\rm z}$ can be measured, we can choose the set of four observables as Pauli operators $\{ Q_j^{(0)} \} = \{ \openone, \sigma^{\rm x}, \sigma^{\rm y}, \sigma^{\rm z} \}$. The operator $\openone$ denotes a trivial measurement, i.e.~the outcome is always $+1$. Measurements of other three Pauli operators can be obtained by applying basis-adjusting operations (Clifford gates) $\{ [\openone], [R_{\rm x}], [R_{\rm z}]^3[R_{\rm x}][R_{\rm z}] \}$ before the measurement of $\sigma^{\rm z}$. Because of the error in the measurement of $\sigma^{\rm z}$ and errors in basis-adjusting operations, the measured observables $\{ Q_j \}$ are not exactly $\{ Q_j^{(0)} \}$. When the overall error is small, observables $\{ Q_j \}$ are still linearly independent. We introduce the matrix $M^{\rm out}_{j, \sigma} = \rbraket{Q_j}{\sigma}$, and $M^{{\rm out} (0)}$ is the matrix corresponding to $\{ Q_j^{(0)} \}$. observables $\{ Q_j \}$ are linearly independent if $M^{\rm out}$ is invertible. We have that $M^{\rm out}$ is always invertible if $\norm{ M^{\rm out} - M^{{\rm out} (0)} }_{\rm max} < \frac{1}{4}s_{\rm min}(M^{{\rm out} (0)}) = \frac{1}{4}$. The initial state without error is decomposed as $\rbra{Q^{(0)}} = \sum_{j=1}^4 q_j \rbra{Q_j}$. Coefficients form a $4$-dimensional row vector $q = [q_1 ~ q_2 ~ q_3 ~ q_4]$. The decomposition is given by $q = \rbra{Q^{(0)}} M^{{\rm out} -1}$.

\section{Quantum gate set tomography}
\label{app:GST}

To measure a set of operations $\{ \bar{\OO}_1, \ldots, \bar{\OO}_N \}$ on $n$ qubits using GST, we need to choose a set of $4^n$ linearly independent initial states $\{ \bar{\rho}_k \}$ and a set of $4^n$ linearly independent observables $\{ \bar{Q}_j \}$. Given these initial states and observables, we measure expected values
\begin{eqnarray}
\tilde{\OO}_{j, k} = \rbra{\bar{Q}_j} \bar{\OO} \rket{\bar{\rho}_k}.
\label{eq:Ot}
\end{eqnarray}
Here, $\bar{\OO}$ is one of operations $\{ \bar{\OO}_1, \ldots, \bar{\OO}_N \}$.

The matrix $\tilde{\OO}$ is equivalent to $\bar{\OO}$ up to a transformation. Because $\bar{\OO}_{\sigma, \tau} = \rbra{\sigma} \bar{\OO} \rket{\tau}$ and $\sum_\sigma \rket{\sigma}\rbra{\sigma} = \openone$ (the sum is taken over all Pauli operators), we have
\begin{eqnarray}
\tilde{\OO} = \bar{M}^{\rm out} \bar{\OO} \bar{M}^{\rm in},
\end{eqnarray}
where $\bar{M}^{\rm in}$ and $\bar{M}^{\rm out}$ are matrices defined as $\bar{M}^{\rm in}_{\sigma, k} = \rbraket{\sigma}{\bar{\rho}_k}$ and $\bar{M}^{\rm out}_{j, \sigma} = \rbraket{\bar{Q}_j}{\sigma}$. We remark that initialisation error and measurement error are included in $\bar{M}^{\rm in}$ and $\bar{M}^{\rm out}$, respectively. We cannot measure matrices $\bar{M}^{\rm in}$ and $\bar{M}^{\rm out}$ independently, therefore we cannot determine $\bar{\OO}$ using GST. By taking $\bar{\OO}$ as the identity operation (i.e.~$\bar{\OO} = \openone$) in Eq.~(\ref{eq:Ot}), we can measure
\begin{eqnarray}
g = \bar{M}^{\rm out} \bar{M}^{\rm in}.
\label{eq:g}
\end{eqnarray}

The estimation of $\bar{\OO}$ is given by
\begin{eqnarray}
\hat{\OO} = T g^{-1} \tilde{\OO} T^{-1} = T \bar{M}^{\rm in -1} \OO \bar{M}^{\rm in} T^{-1}.
\label{eq:Oh}
\end{eqnarray}
Here, $g$ and $\tilde{\OO}$ are obtained by measuring the expected values of observables, and $T$ is an arbitrary invertible matrix. If $\bar{M}^{\rm in}$ and $T$ are different, $\hat{\OO}$ is different from $\bar{\OO}$, but they are always similar matrices. The estimations of states $\rket{\bar{\rho}_k}$ and observables $\rbra{\bar{Q}_j}$ are given by
\begin{eqnarray}
\rket{\hat{\rho}_k} &=& T_{\bullet,k} = T\bar{M}^{{\rm in} -1} \rket{\bar{\rho}_k}, \label{eq:hatrho} \\
\rbra{\hat{Q}_j} &=& (g T^{-1})_{j,\bullet} = \rbra{\bar{Q}_j} \bar{M}^{\rm in} T^{-1}. \label{eq:hatQ}
\end{eqnarray}
Here, $M_{\bullet,k}$ ($M_{j,\bullet}$) denotes the $k^\text{th}$ column ($j^\text{th}$ row) of the matrix $M$.

We introduce matrices $\hat{M}^{\rm in}$ and $\hat{M}^{\rm out}$ defined as $\hat{M}^{\rm in}_{\sigma, k} = \rbraket{\sigma}{\hat{\rho}_k}$ and $\hat{M}^{\rm out}_{j, \sigma} = \rbraket{\hat{Q}_j}{\sigma}$, respectively. Then $\hat{M}^{\rm in} = T$ and $\hat{M}^{\rm out} = g T^{-1}$.

For a sequence of operations $\bar{\OO}_1, \ldots, \bar{\OO}_N$, because $\hat{\OO}_i$ and $\bar{\OO}_i$ are similar matrices up to the same transformation independent of the operation (i.e.~the index $i$), we have
\begin{eqnarray}
\rbra{\bar{Q}_j} \bar{\OO}_N \cdots \bar{\OO}_1 \rket{\bar{\rho}_k} = \rbra{\hat{Q}_j} \hat{\OO}_N \cdots \hat{\OO}_1 \rket{\hat{\rho}_k}.
\end{eqnarray}
Therefore, although estimations $\{ \rket{\hat{\rho}_k}, \rbra{\hat{Q}_j}, \hat{\OO}_i \}$ may be different from their correspondences $\{ \rket{\bar{\rho}_k}, \rbra{\bar{Q}_j}, \bar{\OO}_i \}$, they can always provide the correct prediction for the expected value of an observable in an initial state going through a sequence of operations.

\section{Stability of the quantum gate set tomography}
\label{app:stability}

We define
\begin{eqnarray}
\bar{\varepsilon}_{\rm in} &=& \max\{\norm{\bar{M}^{\rm in}_m - M^{{\rm in}(0)}}_2 \vert m=1,\dots,n\}, \\
\bar{\varepsilon}_{\rm out} &=& \max\{\norm{\bar{M}^{\rm out}_m - M^{{\rm out}(0)}}_2 \vert m=1,\dots,n\}, \\
\bar{\varepsilon}_\OO &=& \norm{\bar{\OO} - \OO^{(0)}}_2,
\end{eqnarray}
which describe severities of the initialisation error, measurement error and operation error, respectively. Here, $\bar{M}^{\rm in}_m$ and $\bar{M}^{\rm out}_m$ are matrices corresponding to the $m^\text{th}$ qubit. The overall matrices of $n$ qubits are $\bar{M}^{\rm in} = \bigotimes_{m=1}^n \bar{M}^{\rm in}_m$ and $\bar{M}^{\rm out} = \bigotimes_{m=1}^n \bar{M}^{\rm out}_m$.

Similar to the analyse of the linear independence of basis operations (i.e.~the invertibility of the matrix $A$, see Appendix~\ref{app:BOth}), we have that $\bar{M}^{\rm in}$ and $\bar{M}^{\rm out}$ are always invertible, i.e.~$g = \bar{M}^{\rm out}\bar{M}^{\rm in}$ is always invertible, if $\bar{\varepsilon}_{\rm in} < s_{\rm min}(M^{{\rm in}(0)})$ and $\bar{\varepsilon}_{\rm out} < s_{\rm min}(M^{{\rm out}(0)})$. Choosing $\{ \rho_k^{(0)} \}$ and $\rbra{Q^{(0)}}$ as in Appendix~\ref{app:decom}, we have $s_{\rm min}(M^{{\rm in}(0)}) = \frac{1}{2}\sqrt{\frac{5-\sqrt{17}}{2}} \simeq 0.3311$ and $s_{\rm min}(M^{{\rm out}(0)}) = 1$.

We choose $T = M^{{\rm in}(0) \otimes n}$, then $\hat{M}^{\rm in} = \bigotimes_{m=1}^n \hat{M}^{\rm in}_m$ and $\hat{M}^{\rm out} = \bigotimes_{m=1}^n \hat{M}^{\rm out}_m$, where $\hat{M}^{\rm in}_m$ and $\hat{M}^{\rm out}_m$ are matrices corresponding to the $m^\text{th}$ qubit. The severity of errors in estimations of initial states is
\begin{eqnarray}
\hat{\varepsilon}_{\rm in} &=& \max\{\norm{\hat{M}^{\rm in}_m - M^{{\rm in}(0)}}_2  \vert m=1,\dots,n\} = 0,
\end{eqnarray}
and the severity of errors in estimations of observables is
\begin{eqnarray}
\hat{\varepsilon}_{\rm out} &=& \max\{\norm{\hat{M}^{\rm out}_m - M^{{\rm out}(0)}}_2  \vert m=1,\dots,n\} \notag \\
&\leq & (\bar{\varepsilon}_{\rm out}\bar{\varepsilon}_{\rm in} + \norm{M^{{\rm in}(0)}}_2\bar{\varepsilon}_{\rm out} + \norm{M^{{\rm out}(0)}}_2\bar{\varepsilon}_{\rm in}) \notag \\
&&\times \norm{M^{{\rm in}(0)-1}}_2.
\end{eqnarray}
Here, we have used that
\begin{eqnarray}
&& \hat{M}^{\rm out}_m - M^{{\rm out}(0)} \notag \\
&=& (\bar{M}^{\rm out}_m\bar{M}^{\rm in}_m - M^{{\rm out}(0)}M^{{\rm in}(0)}) M^{{\rm in}(0)-1} \notag \\
&=& [ (\bar{M}^{\rm out}_m - M^{{\rm out}(0)})(\bar{M}^{\rm in}_m - M^{{\rm in}(0)}) \notag \\
&&+ (\bar{M}^{\rm out}_m - M^{{\rm out}(0)})M^{{\rm in}(0)} \notag \\
&&+ M^{{\rm out}(0)}(\bar{M}^{\rm in}_m - M^{{\rm in}(0)}) ] M^{{\rm in}(0)-1}.
\end{eqnarray}
Choosing $\{ \rho_k^{(0)} \}$ and $\rbra{Q^{(0)}}$ as in Appendix~\ref{app:decom}, we have $\norm{M^{{\rm in}(0)}}_2 = \frac{1}{2}\sqrt{\frac{5+\sqrt{17}}{2}} \simeq 1.0679$, $\norm{M^{{\rm out}(0)}}_2 = 1$ and $\norm{M^{{\rm in}(0)-1}}_2 = s_{\rm min}^{-1}(M^{{\rm in}(0)}) = 2\sqrt{\frac{2}{5-\sqrt{17}}} \simeq 3.0204$.

The severity of the error in the estimation of a $n$-qubit operation is
\begin{eqnarray}
\hat{\varepsilon}_\OO &=& \norm{\hat{\OO} - \OO^{(0)}}_2 \notag \\
&\leq & \norm{\hat{\OO} - \bar{\OO}}_2 + \norm{\bar{\OO} - \OO^{(0)}}_2 \notag \\
&\leq & \frac{2\bar{\varepsilon}_{\rm in}^{(n)}}{[s_{\rm min}(M^{{\rm in}(0)})]^n - \bar{\varepsilon}_{\rm in}^{(n)}} (\norm{\OO^{(0)}}_2 + \bar{\varepsilon}_\OO) \notag \\
&&+ \bar{\varepsilon}_\OO,
\end{eqnarray}
as we will show next. Here,
\begin{eqnarray}
\bar{\varepsilon}_{\rm in}^{(n)} = (\norm{M^{{\rm in}(0)}}_2 + \bar{\varepsilon}_{\rm in})^n - \norm{M^{{\rm in}(0)}}_2^n.
\end{eqnarray}

For an invertible matrix $A$, we have
\begin{eqnarray}
\norm{A^{-1}}_2 &=& \sup_{x\neq 0} \frac{\norm{A^{-1}x}_2}{\norm{x}_2} = \sup_{y\neq 0} \frac{\norm{y}_2}{\norm{Ay}_2}.
\end{eqnarray}
Then, using the inequality~(\ref{ineq_I}), we have
\begin{eqnarray}
\norm{A^{-1}}_2 \leq \frac{1}{s_{\rm min}(A^{(0)}) - \norm{A-A^{(0)}}_2}.
\label{ineq_II}
\end{eqnarray}

We have the expression
\begin{eqnarray}
&& \hat{\OO} - \bar{\OO} \notag \\
&=& M^{{\rm in}(0)\otimes n} \bar{M}^{{\rm in}-1} \bar{\OO} \bar{M}^{\rm in} (M^{{\rm in}(0)\otimes n})^{-1} - \bar{\OO} \notag \\
&=& (M^{{\rm in}(0)\otimes n} - \bar{M}^{\rm in}) \bar{M}^{{\rm in}-1} \bar{\OO} \notag \\
&&\times (\bar{M}^{\rm in} - M^{{\rm in}(0)\otimes n}) (M^{{\rm in}(0)\otimes n})^{-1} \notag \\
&&+ (M^{{\rm in}(0)\otimes n} - \bar{M}^{\rm in}) \bar{M}^{{\rm in}-1} \bar{\OO} \notag \\
&&+ \bar{\OO} (\bar{M}^{\rm in} - M^{{\rm in}(0)\otimes n}) (M^{{\rm in}(0)\otimes n})^{-1}.
\end{eqnarray}
First, we have $\norm{\bar{\OO}}_2 \leq \norm{\OO^{(0)}}_2 + \bar{\varepsilon}_\OO$. Second, using
\begin{eqnarray}
&& \norm{A\otimes B - C\otimes D}_2 \notag \\
&=& \norm{A\otimes B - A\otimes D + A\otimes D - C\otimes D}_2 \notag \\
&\leq & \norm{A}_2 \norm{B - D}_2 + \norm{A - C}_2 \norm{D}_2,
\end{eqnarray}
we have
\begin{eqnarray}
&& \norm{\bar{M}^{\rm in} - M^{{\rm in}(0)\otimes n}}_2 \notag \\
&\leq & \bar{\varepsilon}_{\rm in} \sum_{h = 1}^n \norm{M^{{\rm in}(0)}}_2^{n-h} \prod_{m=1}^{h-1} \norm{\bar{M}^{\rm in}_m}_2 \notag \\
&\leq & \bar{\varepsilon}_{\rm in} \sum_{h = 1}^n \norm{M^{{\rm in}(0)}}_2^{n-h} (\norm{M^{{\rm in}(0)}}_2 + \bar{\varepsilon}_{\rm in})^{h-1} \notag \\
&=& \bar{\varepsilon}_{\rm in}^{(n)}.
\end{eqnarray}
Third, using the inequality~(\ref{ineq_II}), we have
\begin{eqnarray}
&& \norm{\bar{M}^{{\rm in}-1}}_2 \leq \frac{1}{[s_{\rm min}(M^{{\rm in}(0)})]^n - \bar{\varepsilon}_{\rm in}^{(n)}}.
\end{eqnarray}

We remark that for a $d$-dimensional matrix $M$, $\norm{M}_{\rm max} \leq \norm{M}_2 \leq d\norm{M}_{\rm max}$.

\section{Upper bound of the cost}
\label{app:upper}

We consider compensation method and take $\lambda = 1$, i.e.~the $n$-qubit operation without error is realised as $\OO^{(0)} = \OO + \EE$, where $\EE$ is decomposed using basis operations as shown in Eq.~\ref{eq:decomE}. Then the cost for correcting the error in $\OO$ is determined by
\begin{eqnarray}
C_\OO = 1 + \sum_{i_1,\ldots,i_n} \abs{q_{i_1,\ldots,i_n}}.
\end{eqnarray}
Decomposition coefficients are determined by $q = (A_1 \otimes\cdots\otimes A_n)^{-1} E$, where $q$ and $E$ are defined in Eqs.~(\ref{eq:q},\ref{eq:E}). Here, $q$ and $E$ are $16^n$-dimensional vectors, and $A_1 \otimes\cdots\otimes A_n$ is a $16^n$-dimensional matrix. Therefore, for each element of $q$,
\begin{eqnarray}
\abs{q_{i_1,\ldots,i_n}} \leq 16^n \norm{(A_1 \otimes\cdots\otimes A_n)^{-1}}_{\rm max} \norm{\EE}_{\rm max}.
\end{eqnarray}
Here, we have used that the maximum absolute value of an element of $E$ is $\norm{\EE}_{\rm max}$. Because $\EE = \OO^{(0)} - \OO$, we have $\norm{\EE}_{\rm max} = \epsilon_\OO$. Using the inequality~(\ref{ineq_II}), we have
\begin{eqnarray}
&& \norm{(A_1 \otimes\cdots\otimes A_n)^{-1}}_{\rm max} \notag \\
&=& \prod_{l=1}^n \norm{A_l^{-1}}_{\rm max}
\leq \prod_{l=1}^n \norm{A_l^{-1}}_2 \notag \\
&\leq & \prod_{l=1}^n \frac{1}{s_{\rm min}(A^{(0)})-\norm{A_l-A^{(0)}}_2} \notag \\
&\leq & \prod_{l=1}^n \frac{1}{s_{\rm min}(A^{(0)})-16\norm{A_l-A^{(0)}}_{\rm max}} \notag \\
&\leq & \frac{1}{[s_{\rm min}(A^{(0)})-16\epsilon_{\rm max}]^n}.
\end{eqnarray}
Here, $\epsilon_{\rm max} = \max\{ \norm{A_l-A^{(0)}}_{\rm max} \vert l = 1,\ldots,n \}$. There are total $16^n$ decomposition coefficients, therefore
\begin{eqnarray}
C_\OO \leq 1 + \frac{16^{2n}\epsilon_\OO}{(s_{\rm min}(A^{(0)})-16\epsilon_{\rm max})^n}.
\end{eqnarray}
Here, we have assumed that $\epsilon_{\rm max} < s_{\rm min}(A^{(0)})/16$.

We consider using the set of initial states with errors $\{ \rket{\rho_k} \}$ to realise the initial state $\rket{\rho_{k_0}^{(0)}}$, which is in the set of initial states without error $\{ \rket{\rho_k^{(0)}} \}$. The initial state can be decomposed as $\rket{\rho_{k_0}^{(0)}} = \rket{\rho_{k_0}} + \sum_k (q_k-\delta_{k,k_0})\rket{\rho_k}$ (see Appendix~\ref{app:decom}). We use $\epsilon_{\rm in} = \norm{M^{\rm in} - M^{{\rm in}(0)}}_{\rm max}$ as the measure of the error severity in initial states. Using $\norm{\rket{\rho_{k_0}} - \rket{\rho_{k_0}^{(0)}}}_{\rm max} \leq \epsilon_{\rm in}$ and $\norm{M^{{\rm in}-1}}_{\rm max} \leq 4[s_{\rm min}(M^{{\rm in}(0)}) - \epsilon_{\rm in}]^{-1}$, we have the cost for correcting errors in initial states
\begin{eqnarray}
C_{\rm in} &=& \sum_k \abs{q_k} \leq 1 + \sum_k \abs{q_k - \delta_{k,k_0}} \notag \\
&\leq & 1 + \frac{4^2\epsilon_{\rm in}}{s_{\rm min}(M^{{\rm in}(0)}) - \epsilon_{\rm in}}.
\end{eqnarray}

It is similar for observables. We consider using the set of observables with errors $\{ \rbra{Q_j} \}$ to realise the observable $\rbra{Q_{j_0}^{(0)}}$, which is in the set of observables without error $\{ \rbra{Q_j^{(0)}} \}$. We use $\epsilon_{\rm out} = \norm{M^{\rm out} - M^{{\rm out}(0)}}_{\rm max}$ as the measure of the error severity in observables. Then, the cost for correcting errors in measured observables is
\begin{eqnarray}
C_{\rm out} \leq 1 + \frac{4^2\epsilon_{\rm out}}{s_{\rm min}(M^{{\rm out}(0)}) - \epsilon_{\rm out}}.
\end{eqnarray}

\section{Error models}
\label{app:model}

We consider a quantum computer with the following operations. The initialisation $\II^{(0)} = [\pi] + [\pi \sigma^{\rm x}]$, which prepares the state $\ket{0}$. The projective measurement $[\pi]$. Single-qubit Clifford gates $[R_{\rm x}]$ and $[R_{\rm z}]$. The single-qubit non-Clifford gate $[T]$, where $T = \openone \cos\frac{\pi}{8} + i\sigma^{\rm x}\sin\frac{\pi}{8}$. Two-qubit maximally entangling gate $[\Lambda]$, where $\Lambda = \frac{1}{\sqrt{2}}(\openone + i\sigma^{\rm z}\otimes\sigma^{\rm z})$, which is equivalent to the controlled-{\small NOT} gate and controlled-phase gate up to single-qubit gates. The sixteen basis operations can be realised as shown in Table~\ref{tab:bases}. In order to perform GST, we choose initial states and observables as in Appendix~\ref{app:decom}, and we choose the invertible matrix $T = M^{{\rm in}(0) \otimes n}$ for $n$ qubits.

For the initialisation, the state prepared is $\rho_0$ rather than $\ketbra{0}{0}$. We can always express the initialisation operation with error as $\II = \NN_{\rm i} \II$, where $\NN_{\rm i}(\ketbra{0}{0}) = \rho_0$.

A POVM is defined by a set of operators $\{ E_k \}$ satisfying $\sum_k E_k^\dag E_k = \openone$. In a POVM, the state is mapped to $E_k \rho E_k^\dag$ when the outcome is $k$. When the measurement has error, we may not be able to obtain all the information $k$. Usually there are only two outcomes corresponding to $\ket{0}$ and $\ket{1}$, respectively. In this case, maybe several $k$ values correspond to the same outcome $\nu = 0,1$. Therefore, we model the projective measurement $[\pi]$ with error as $\MM \rho = \sum_{k\in K_0} E_k \rho E_k^\dag$, where $K_\nu$ is the set of $k$ corresponding to the measurement outcome $\nu$.

For a gate without error $\GG^{(0)}$, the gate with error can be expressed as $\GG = \NN_{\rm a} \GG^{(0)} \NN_{\rm b}$. Any noisy gate can be expressed in this form: Because $\GG^{(0)}$ is invertible, we can always take $\NN_{\rm b} = [\openone]$ and $\NN_{\rm a} = \GG \GG^{(0)-1}$.

We suppose that time costs of the measurement $[\pi]$ and the two-qubit gate $[\Lambda]$ are the same, and time costs of single qubit gates are negligible.

We distinguish the identity operation and the memory operation. Without error, both of them are the same operation $[\openone]$. In any case, the identity operation is $[\openone]$, which means that the next operation is performed immediately, so it takes no time and there is not any memory error. When the memory operation is performed, the qubit waits for the next operation, so memory errors may occur on it. We apply the identity operation for measuring the matrix $g$ (see Appendix~\ref{app:GST}). In the basis operation set, the operation $[\openone]$ is replaced by the memory operation.

We set the cycle time of the computing as the time cost of the measurement and the two-qubit gate. In one cycle, only one operation is performed on a qubit. If the operation is a single-qubit gate, the gate is performed at the middle of the cycle, i.e.~the overall operation is $\NN_{\rm m} \GG \NN_{\rm m}$, where $\NN_{\rm m}$ denotes memory noise. If no gate or measurement is performed in the cycle, the overall operation is $\NN_{\rm m}^2$, which is the error version of the operation $[\openone]$ in the basis operation set.

We suppose that the single-qubit noise is described by the superoperator $\EE^{(1)}(\epsilon)$, and the two-qubit noise is described by the superoperator $\EE^{(2)}(\epsilon)$. Here, $\epsilon$ is a parameter describing the intensity of the noise. Then, the initialisation noise is $\NN_{\rm i} = \EE^{(1)}(\frac{\epsilon}{10})$, and the measurement with noise is $\tilde{M} = \mathcal{E}^{(1)}(\frac{\epsilon}{2})[\pi]\EE^{(1)}(\frac{\epsilon}{2})$. For single-qubit gates, $\NN_{\rm a} = \NN_{\rm b} = \EE^{(1)}(\frac{\epsilon}{20})$. For the two-qubit gate, $\NN_{\rm a} = \NN_{\rm b} = \EE^{(2)}(\frac{\epsilon}{2})$. For the memory operation, $\NN_{\rm m} = \EE^{(1)}(\frac{\epsilon}{200})$.

\subsection{Depolarising Error}

The single-qubit depolarising noise is
\begin{eqnarray}
\EE^{(1)}(\epsilon) = (1-\frac{4\epsilon}{3}) [\openone]
+ \frac{\epsilon}{3} \sum_{\alpha = 0}^3 [\sigma^\alpha],
\end{eqnarray}
where $\sigma^0$, $\sigma^1$, $\sigma^2$ and $\sigma^3$ correspond to $\openone$, $\sigma^{\rm x}$, $\sigma^{\rm y}$ and $\sigma^{\rm z}$, respectively. The two-qubit depolarising noise is
\begin{eqnarray}
\EE^{(2)}(\epsilon) = (1-\frac{16\epsilon}{15}) [\openone \otimes \openone]
+ \frac{\epsilon}{15} \sum_{\alpha,\beta = 0}^3 [\sigma^\alpha \otimes \sigma^\beta].
\end{eqnarray}
The x-axis (error rate) in Fig.~\ref{fig:cost}(a) is $\epsilon$ of the two-qubit gate.

\subsection{Dephasing Error}

The single-qubit dephasing noise is
\begin{eqnarray}
\EE^{(1)}(\epsilon) = (1-\epsilon) [\openone]
+ \epsilon [\sigma^{\rm z}].
\end{eqnarray}
The two-qubit dephasing noise is
\begin{eqnarray}
\EE^{(2)}(\epsilon) &=& (1-\epsilon) [\openone \otimes \openone]
+ \frac{\epsilon}{3}( [\openone \otimes \sigma^{\rm z}] \notag \\
&&+  [\sigma^{\rm z} \otimes \openone] +  [\sigma^{\rm z} \otimes \sigma^{\rm z}]).
\end{eqnarray}

\subsection{Damping Error}

The single-qubit damping noise is
\begin{eqnarray}
\EE^{(1)}(\epsilon) &=& [\frac{\openone+\sigma^{\rm z}}{2} + \sqrt{1-\epsilon}\frac{\openone-\sigma^{\rm z}}{2}] \notag \\
&&+ [\sqrt{\epsilon}\frac{\sigma^{\rm x}+i\sigma^{\rm y}}{2}].
\end{eqnarray}
The two-qubit damping noise is
\begin{eqnarray}
\EE^{(2)}(\epsilon) = \mathcal{E}^{(1)}(\frac{\epsilon}{2})
\otimes \mathcal{E}^{(1)}(\frac{\epsilon}{2}).
\end{eqnarray}

\subsection{Over-rotation error}

Noise is gate dependent. Initialisation, measurement and memory operation are perfect, i.e.~$\EE^{(1)} = [\openone]$ for these operations. Only gates have noise. For gate $R_{\rm x}$, $\EE^{(1)}(\epsilon) = [\openone\cos\frac{\epsilon\pi}{4} + i\sigma^{\rm x}\sin\frac{\epsilon\pi}{4}]$. For gate $R_{\rm x}$, $\EE^{(1)}(\epsilon) = [\openone\cos\frac{\epsilon\pi}{4} + i\sigma^{\rm z}\sin\frac{\epsilon\pi}{4}]$. For gate $T$, $\EE^{(1)}(\epsilon) = [\openone\cos\frac{\epsilon\pi}{8} + i\sigma^{\rm z}\sin\frac{\epsilon\pi}{8}]$. For gate $\Lambda$, $\EE^{(2)}(\epsilon) = [\openone\otimes\openone\cos\frac{\epsilon\pi}{4} + i\sigma^{\rm z}\otimes\sigma^{\rm z}\sin\frac{\epsilon\pi}{4}]$. The x-axis (over rotation) in Fig.~\ref{fig:cost}(a) is $\epsilon$ of the two-qubit gate.

\subsection{Random-field error}

Noise is gate dependent. For each operation, the noise $\EE^{(1,2)}(\epsilon) = [e^{-i\epsilon \pi H}]$ is determined by a Hamiltonian. Here, $H = (h+h^\dag)/2$, and each element of $h$ is randomly generated with a uniform distribution in the unit circle. We remark that the noise is time independent, i.e.~the noise is the same for the same gate implemented at different times.

\subsection{Random-operation error}

The operation without noise is $\GG^{(0)}$. The operation with noise is $\GG(\epsilon)$, which depends on the error parameter. As the same as other models, the error parameter $\epsilon$ is operation dependent. Each operation can be expressed using a $\chi$-matrix~\cite{Nielsen2010}. We suppose the $\chi$-matrix corresponding to $\GG(\epsilon)$ is $\chi$, and the $\chi$-matrix corresponding to $\GG^{(0)}$ is $\chi^{(0)}$. To generate $\chi$, firstly we generate a Hermitian matrix around $\chi^{(0)}$, which is $\chi' = \chi^{(0)} + \epsilon H$, where $H$ is generate as the same as the random Hamiltonian. Second, if $\GG^{(0)}$ is not measurement, $\GG(\epsilon)$ should be trace preserving. However, $\chi'$ may correspond to a non-trace preserving operation. In this case, we project $\chi'$ to the subspace in the matrix space that corresponds to trace preserving operations, i.e.~$\chi''$ is the matrix closest to $\chi'$ and corresponds to a trace preserving operation. If $\GG^{(0)}$ is measurement, $\chi'' = \chi'$. Third, $\chi''$ may not be positive semi-definite. Therefore, we take $\chi''' = \chi''+\lambda_{\rm min}\openone$ if the minimum eigenvalue $\lambda_{\rm min}$ of $\chi''$ is negative; otherwise $\chi''' = \chi''$. Finally, $\chi = f\chi'''$, where the factor $f$ makes sure that the operation is still trace preserving and the maximum eigenvalue of $\chi$ is smaller than $1$.

\section{Instruction of the implementation of the quasi-probability method}
\label{app:instruction}

\RED{This section is a self-contained description of how to implement QEM using the quasi-probability decomposition. There are three steps: first, implement GST; second, compute the quasi-probability decomposition; third, implement the quasi-probability decomposition using the Monte Carlo approach.
 
\subsection{Implementation of gate set tomography}

General discussions of GST are given in the main text and Appendix~\ref{app:GST}, therefore, here we describe GST in a more concrete way. GST is implemented to measure all gates used in the quantum computation. We discuss how to measure single qubit gates first and two-qubit gates afterwards.

To measure a single-qubit gate using GST, we prepare initial states $\ket{0}$, $\ket{1}$, $\ket{+}$, and $\ket{y+}$, where $\ket{+}$, and $\ket{y+}$ are the eigenstates of Pauli operators $\sigma^{\rm x}$ and $\sigma^{\rm y}$ with the eigenvalue $+1$, respectively. We denote these states $\bar{\rho}_1$,  $\bar{\rho}_2$, $\bar{\rho}_3$ and $\bar{\rho}_4$, respectively. These initial states can be noisy states, which is the essential advantage of GST, i.e.~GST can tolerate state preparation and measurement errors. Then, we apply the gate that we want to measure, for instance, Hadamard gate, $T$ gate and $T^\dag$ gate in the {\small SWAP}-test circuit. Here we will use $\bar{\OO}$ (superoperator acting on a reduced density matrix) to denote the gate to be measured, which has noise. Subsequently, we measure expectation values for four observables, $\openone$, $\sigma^{\rm x}$, $\sigma^{\rm y}$ and $\sigma^{\rm z}$, respectively. Here $\openone$ is a trivial observable whose measurement outcome is always $+1$. We denote these observables as $\bar{Q}_1$, $\bar{Q}_2$, $\bar{Q}_3$, $\bar{Q}_4$, and measurements of these observables can also be noisy. Then, by repeating the experiment to compute the mean value of observables, we can construct the $4\times 4$ matrix $\tilde{\OO}$, and matrix elements are
\begin{eqnarray}
\tilde{\OO}_{j,k} = \Tr \left( \bar{Q}_j \bar{\OO} \bar{\rho}_k \right).
\end{eqnarray}
Similarly, we can obtain the $4\times 4$ matrix $g$ by choosing not to apply any gate to initial state, so that the matrix elements are
\begin{eqnarray}
g_{j,k} = \Tr \left( \bar{Q}_j \bar{\rho}_k \right).
\end{eqnarray}
This process is implemented for each qubit and each type of single-qubit gate (including all basis operations).

One will find that there is a freedom in the specification of the gate $\bar{\OO}$. Legitimate variants can be obtained as
\begin{eqnarray}
\hat{\OO} = T g^{-1} \tilde{\OO} T^{-1},
\end{eqnarray}
where $T$ is an invertible $4\times 4$ matrix. The matrix $T$ can be different for different qubits but must be the same for all gates on the same qubit. We can choose $T$ to minimise the cost in QEM. In the case that the error rate of preparing initial states is low, we can take
\begin{eqnarray}
T = \begin{pmatrix}
1&1&1&1 \\
0&0&1&0 \\
0&0&0&1 \\
1&-1&0&0 \\
\end{pmatrix},
\end{eqnarray}
which approximately minimise the cost according to our experience.

Estimations of the initial state $\bar{\rho}_k$ and the observable $\bar{Q}_j$ are respectively
\begin{eqnarray}
\rket{\hat{\rho}_k} &=& T_{\bullet,k}, \\
\rbra{\hat{Q}_j} &=& (g T^{-1})_{j,\bullet}.
\end{eqnarray}
Here, $M_{\bullet,k}$ ($M_{j,\bullet}$) denotes the $k^\text{th}$ column ($j^\text{th}$ row) of the matrix $M$.

To measure a two-qubit gate using GST, the procedure is basically the same. The only difference is that there are $16$ initial states and $16$ observables to be measured. Initial states are the tensor products of single-qubit initial states, i.e.~$\bar{\rho}^{(1)}_{k_1} \otimes \bar{\rho}^{(2)}_{k_2}$, and observables are tensor products of single-qubit observables, i.e.~$\bar{Q}^{(1)}_{j_1} \otimes \bar{Q}^{(2)}_{j_2}$. Here, the superscript is the label of the qubit. Accordingly, the matrix $g = g^{(1)}\otimes g^{(2)}$, which is the tensor product of $g$ matrices of two qubits, and similarly the matrix $T = T^{(1)}\otimes T^{(2)}$, which is the tensor product of $T$ matrices of two qubits. We need to implement two-qubit gate GST for each pair of qubits that the two-qubit gate may be performed on.

\subsection{Quasi-probability decomposition}

Using results obtained from GST, we can compute the quasi-probability decomposition. From GST we obtain estimations of initial states, observables to be measured, gates (including basis operations), and they are
\begin{eqnarray}
\rket{\hat{\rho}_k} &~:~& \rm{Initial~state} \notag \\
\rbra{\hat{Q}_j}  &~:~& \rm{Observable} \notag \\
\hat{\OO} &~:~& \rm{Gate} \notag \\
\hat{\BB}_i &~:~& \rm{Basis~operation} \notag
\end{eqnarray}
These estimations are utilised to compute the quasi-probability decomposition.

Now, we focus on the inverse method. We use $\OO^{(0)}$ to denote the Pauli transfer matrix of the ideal gate (without error). The estimation of the Pauli transfer matrix of the actual gate with noise (i.e.~$\bar{\OO}$) is $\hat{\OO}$, which is obtained in GST. To compute the decomposition, first, we compute the ideal matrix $\OO^{(0)}$; second, we compute the inverse of the noise
\begin{eqnarray}
\NN^{-1} = \OO^{(0)} \hat{\OO}^{-1};
\end{eqnarray}
and finally, we solve the equation (for single-qubit gate)
\begin{eqnarray}
\NN^{-1} = \sum_i q_{\OO,i} \hat{\BB}_i
\end{eqnarray}
to determine quasi-probabilities $q_{\OO,i}$ of the gate $\OO$. We need to compute quasi-probabilities for each qubit and each gate. For example, for the {\small SWAP}-test circuit, we need to compute the decomposition for Hadamard gate, $T$ gate, $T^\dag$ gate of each qubit and controlled-{\small NOT} gate of each pair of qubits that the controlled-{\small NOT} gate may be performed on.

For two-qubit gates, the procedure is the same but tensor products of single-qubit basis operations, i.e.~$\hat{\BB}^{(1)}_{i_1} \otimes \hat{\BB}^{(2)}_{i_2}$ (where the superscribe is the label of the qubit), are used to decompose the inverse of the noise $\NN^{-1}$.

In order to mitigate errors in initial states and measurements of observables, {we should solve the following equations for the quantities $q^{(m)}_{\rho,k}$ and $q^{(m)}_{Q,j}$}:
\begin{eqnarray}
\rket{\rho^{(0)}} &=& \sum_k q^{(m)}_{\rho,k} \rket{\hat{\rho}_k}, \\
\rbra{Q^{(0)}} &=& \sum_k q^{(m)}_{Q,j} \rbra{\hat{Q}_j}
\end{eqnarray}
for each qubit. Here, $m$ is the label of the qubit, $\rket{\rho^{(0)}}$ is the column vector representing the ideal initial state $\ketbra{0}{0}$, and $\rbra{Q^{(0)}}$ is the row vector representing the ideal observable $\sigma^{\rm z}$. 

Before implementing the quasi-probability decomposition on a quantum computer, we compute
\begin{eqnarray}
C^{(m)}_\rho &=& \sum_k \abs{q^{(m)}_{\rho,k}}, \\
C^{(m)}_Q &=& \sum_k \abs{q^{(m)}_{Q,j}}
\end{eqnarray}
for each qubit and
\begin{eqnarray}
C_\OO &=& \sum_i \abs{q_{\OO,i}}
\end{eqnarray}
for each gate.

\subsection{Monte Carlo implementation of the quasi-probability decomposition}

It is vital to note that we use estimations $\hat{\BB}_i$ to decompose the inverse of the noise, but usually there is a difference between $\hat{\BB}_i$ and the actual basis operation $\bar{\BB}_i$. This difference does not cause any error in the final computing result, because the computing result is invariant under a similarity transformation as we have explained in the main text.

Now, we describe how to implement the quasi-probability decomposition on a quantum computer. We suppose the circuit is sequentially performing gates $\OO_1$, $\OO_2$, \ldots, $\OO_N$ on the initial state $\ket{00\ldots 0}$, and the first qubit is measured in the $\sigma^{\rm z}$ basis to read the computing result. The procedure can be generalised to the case of measuring multiple qubits.

First, we generate {a set of random integers: for each qubit $m$, we randomly select an integer $k_m$ such that each integer would be selected with corresponding probability $\abs{q^{(m)}_{\rho,k_m}}/C^{(m)}_\rho$; similarly for each gate $l$, we generate random integer $i_l$ with corresponding probability $\abs{q_{\OO_l,i_l}}/C_{\OO_l}$; and finally we generate random integer $j_1$ with the probability $\abs{q^{(1)}_{Q,j_1}}/C^{(1)}_Q$.}

Second, on the quantum computer, we implement the following quantum computing for once: we initialise the qubit $m$ in the state $\bar{\rho}^{(m)}_{k_m}$; then we sequentially perform gates $\bar{\OO}_1$, $\bar{B}_{i_1}$, $\bar{\OO}_2$, $\bar{B}_{i_2}$, \ldots, $\bar{\OO}_N$, $\bar{B}_{i_N}$; finally we measure the observable $\bar{Q}_{j_1}$. The measurement outcome is $\mu$.

Third, we compute the effective measurement outcome
\begin{eqnarray}
\mu_{\rm eff} = {\rm sgn} \left( \prod_m q^{(m)}_{\rho,k_m} \prod_l q_{\OO_l,i_l} q^{(1)}_{Q,j_1} \right) \mu.
\end{eqnarray}

By repeating these three steps, we can obtain the mean of effective outcomes ${\rm E}[\mu_{\rm eff}]$. The final computing result is $C {\rm E}[\mu_{\rm eff}]$, where
\begin{eqnarray}
C = \prod_m C^{(m)}_\rho \prod_l C_{\OO_l} C^{(1)}_Q.
\end{eqnarray}
}

\section{Error component of Pauli error and leakage error}
\label{app:errorCom}

Taking $\epsilon = p_{\rm x} + p_{\rm y} + p_{\rm z}$, it is obvious that in the inhomogeneous Pauli error model, the error component does not depend on the error rate, i.e.~$\mathcal{E}' = \epsilon^{-1}(p_{\rm x}[\sigma^{\rm x}] + p_{\rm y}[\sigma^{\rm y}] + p_{\rm z}[\sigma^{\rm z}])$. We remark that ratios $p_\alpha/\epsilon$ do not change with $\epsilon$.

For the leakage error model, we take $\epsilon = p$. Then the error component is
\begin{eqnarray}
\mathcal{E}'(\rho) &=& \pi_0 \rho \pi_0 + \frac{\sqrt{1-p} - (1-p)}{p} (\pi_0 \rho \pi_1 + \pi_1 \rho \pi_0) \notag \\
&=& \pi_0 \rho \pi_0 + [1/2+\mathcal{O}(p)] (\pi_0 \rho \pi_1 + \pi_1 \rho \pi_0),
\end{eqnarray}
where $\pi_0 = \ketbra{0}{0}$ and $\pi_1 = \ketbra{1}{1}$. Therefore, $\mathcal{E}'$ varies slowly with $p$.

\end{document}